\documentclass{article}
\usepackage{iclr2027_conference,times}

\usepackage{amsmath,amsfonts,bm}

\def\eqref#1{equation~\ref{#1}}

\def\1{\bm{1}}

\DeclareMathAlphabet{\mathsfit}{\encodingdefault}{\sfdefault}{m}{sl}
\SetMathAlphabet{\mathsfit}{bold}{\encodingdefault}{\sfdefault}{bx}{n}

\usepackage{hyperref}
\usepackage{url}
\usepackage[utf8]{inputenc} 
\usepackage[T1]{fontenc}    
\usepackage{booktabs}       
\usepackage{amsfonts}       
\usepackage{nicefrac}       
\usepackage{microtype}      
\usepackage{xcolor}         
\usepackage{graphicx} 
\usepackage{amssymb}
\usepackage{pifont}
\newcommand{\cmark}{\ding{51}}%
\newcommand{\xmark}{\ding{55}}%

\title{CrossEdit: Cross-Modal Training Enables  Rich Audio-Visual Editing}

\author{%
  {William Chen$^{1\dagger,2}$}\thanks{Equal Contribution. $\dagger$ {Work done while at Adobe.}},  
  Prem Seetharaman$^{1*}$, Ke Chen$^{1*}$, Oriol Nieto$^{1*}$,\\ 
  \textbf{Kevin Duarte$^{1}$, Siddharth Srinivasan Iyer$^{1}$, Mamshad Nayeem Rizve$^{1}$, Zhiwen Cao$^{1}$} \\
  \textbf{Shinji Watanabe$^{2}$, Yuanjun Xiong$^{1}$, Jianming Zhang$^{1}$, Zeyu Jin$^{1}$, Justin Salamon$^{1}$} \\
  Adobe Research$^{1}$ \\
  Carnegie Mellon University$^{2}$ \\
  \texttt{williamchen@cmu.edu, seethara@adobe.com} \\
}

\iclrfinalcopy 
\begin{document}

\maketitle

\begin{abstract}
Multimedia editing requires generative models to interpret complex, compositional instructions and modify only the desired elements across modalities, a capability largely beyond existing systems. Current approaches are typically limited to simple, single-attribute edits within one modality, since diverse, high-quality editing pairs are hard to obtain, especially for cross-modal tasks where aligned audiovisual (AV) data is scarce. We present \textbf{CrossEdit}, a unified omni-modal editing model for images, audio, and video that performs AV movie scene edits zero-shot through cross-modal transfer. Our key observation is that complex instruction-following, once learned in any modality, generalizes to others. We exploit the additive nature of acoustic signals to procedurally generate large-scale audio editing pairs with compositional instructions, and show that fine-tuning on this synthetic data and self-supervised AV masked reconstruction, alongside a targeted set of curated cross-modal tasks, induces instruction-following that transfers zero-shot to unseen modality and instruction combinations. To evaluate this capability, we release \textbf{CrossEditBench}, a human-annotated benchmark of AV edits on movie scenes, and propose \textbf{AV-FES}, a metric that jointly scores instruction following and consistency. We show that our proposed techniques improve performance on zero-shot AV movie scene editing, while maintaining or improving performance on lip-synced speech editing and standard image, video, and audio editing benchmarks. Demos are available at \url{https://wanchichen.github.io/crossedit/}.
\end{abstract}

\section{Introduction}

Recent advances in multi-modal generative models have enabled the synthesis of images, video, and audio from natural language descriptions \citep{valle, bagpiper, chen2026audiochat, bagel, zhou2025transfusion, shi2024lmfusion, stableaudioopen, hung2024tangoflux}, with rapidly improving fidelity and controllability. Yet generation is only half the story. In practice, it is unlikely that a generative model will produce exactly what a user desires in a single attempt — natural language is inherently ambiguous, and creative workflows are inherently iterative. Users will inevitably wish to modify some aspects of generated media while preserving others. This is most difficult for audio-visual (AV) content, where an edit such as adding a motorbike to a street scene must change both picture and sound, keep them synchronized, and leave everything else untouched.

Existing editing models, however, remain largely siloed by modality \citep{chen2026audiochat, wang2023audit, recomposer, lan2025guiding, peng-etal-2024-voicecraft}, and the few multi-modal systems support only simple, single-attribute edits (e.g., ``make her hair blue''). The bottleneck is data: editing pairs are easy to construct for atomic, single-modality edits, but AV-to-AV (AV2AV) pairs with compositional instructions rarely occur naturally and are expensive to curate.

\begin{figure}
    \centering
    \includegraphics[width=0.75\linewidth]{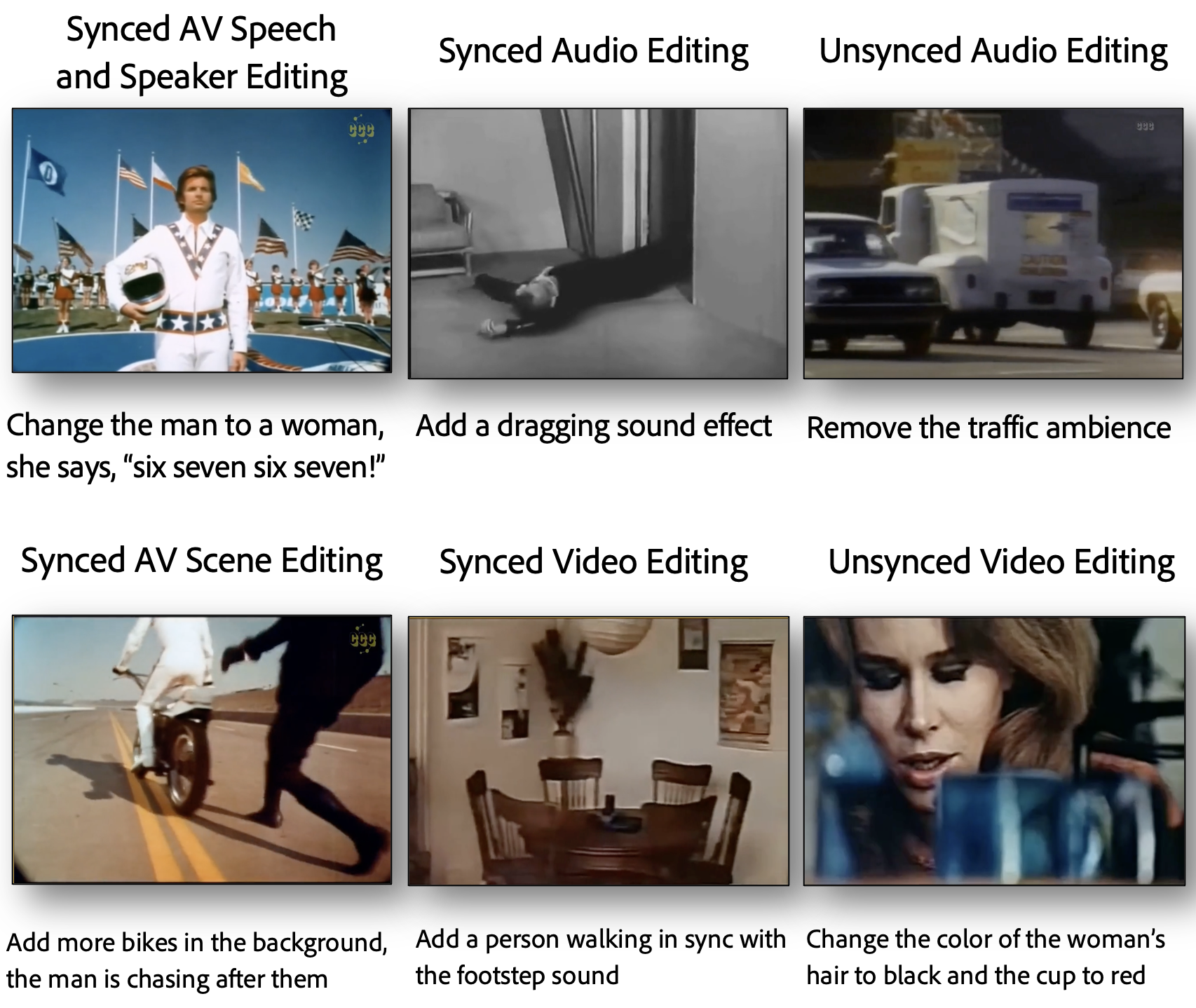}
    \caption{\textbf{Instruction types in CrossEditBench that CrossEdit must handle zero-shot.} Models must implicitly determine which modalities to edit and what to synchronize across audio and video.}
    \label{fig:demo}
\vspace{-0.35cm}
\end{figure}

We present \textbf{CrossEdit}, an omni-modal editing model that can edit real movie scenes from free-form instructions (Figure~\ref{fig:demo}) \textit{zero-shot}, without ever being trained on such data. Our work rests on two hypotheses: (1) \textit{complex instruction-following is modality-agnostic}, so once learned in one modality it transfers to others; and (2) \textit{masked AV reconstruction can serve as a self-supervised AV2AV editing task}. We operationalize the first hypothesis by exploiting an asymmetry between auditory and visual signals. Acoustic signals are physically additive (mixing waveforms yields a valid audio scene), so large-scale pairs of complex instructions and simulated mixtures can be easily generated, whereas visual compositing lacks this physical fidelity. We hypothesize that these synthetic audio editing pairs, combined with a core set of curated cross-modal tasks (such as masked AV reconstruction), can better generalize complex instruction-following capabilities across modalities.

To evaluate zero-shot AV2AV movie scene editing, we introduce CrossEditBench, a human-annotated benchmark of edits on real movie scenes that require jointly reasoning over the instruction, video, and audio, and AV-FES, a novel metric that directly evaluates editing performance with a single interpretable score. Our results show that our proposed techniques allow the model to better perform AV2AV movie scene editing without any training data for the task, while maintaining, if not improving, performance on trained tasks such as video and image editing.

\vspace{-0.2cm}
\section{Related Work}

\paragraph{Multi-modal foundation models.} Unifying images \citep{zhai2022scaling, ho2020denoising, song2021denoising, salimans2022progressive}, video \citep{diffusion_forcing}, audio \citep{whisper, pratap2023scaling, chen-etal-2024-towards-robust, chen2025owls, valle2025fugatto}, and text \citep{gpt3, gpto1, yang2025qwen3} within a single model has drawn significant attention \citep{team2025gemma3, bai2025qwen3, chu2024qwen2, tian2025ualm, chen2026audiochat, bagel, zhou2025transfusion}. Most work pairs visual or audio generation with text \citep{tian2025ualm, bagel, zhou2025transfusion, opuslmv, bagpiper, liao2025mogao}, and any-to-any models \citep{luo2026nextomni,tang2024codi, li2025omniflow} typically cannot output multiple modalities simultaneously. Seedance 2.0 \citep{seedance2026seedance}, LTX-2~\citep{hacohen2026ltx}, and UniAVGen~\citep{zhang2025uniavgen} achieve synchronized AV synthesis via dual-stream diffusion transformers, but can only \textit{generate}, not \textit{edit}.

\paragraph{Instruction-based media editing.} Editing has been explored within each modality: for images, via synthetic editing triplets~\citep{brooks2023instructpix2pix, zhang2023magicbrush, geng2024instructdiffusion, zhao2024ultraedit, wei2025omniedit}; for video, via temporal consistency mechanisms on image diffusion models \citep{wu2023tune, geyer2023tokenflow, qi2023fatezero} and, more recently, large editing datasets \citep{Wu_2025_ICCV,bai2025ditto,he2025openve,wei2025univideo,ju2026editverse}; and for audio, via mixtures of real-world signals for a preset list of supported operations \citep{wang2023audit, recomposer, huang2023makeanaudio}, later extended with LLM-generated instructions \citep{lan2025guiding, chen2026audiochat}. Early AV2AV works relied on cascaded pipelines \citep{ishii2025coherent, polyak2024moviegen, zheng2025avi} or sampling tricks applied to uni-modal models \citep{aved}. End-to-end AV editing has only emerged recently, but concurrent works target narrower settings. For example, SpongeBob \citep{liang2026spongebob} and InstructAV2AV \citep{zheng2026instructav2av} only perform simpler instructions that are doable via inpainting, such as addition and removal. Spongebob requires per-modality captions, an object mask, and a reference image, so the user rather than the model decides what changes in each modality; its code, weights, and benchmark are also unreleased. Both train on paired AV editing data for their target tasks and are AV-only, whereas CrossEdit is omni-modal and edits real movie scenes from free-form instructions zero-shot. Most recently, MiniMax H3 \citep{minimax2026h3} unifies generation and editing of images, audio, and video within a single model. To our knowledge, CrossEdit is the first work to show that instruction-based AV2AV scene editing can be acquired zero-shot through cross-modal transfer of instruction following.

\begin{table}[t]
    \centering
    \caption{\textbf{Training and Evaluation tasks for CrossEdit.} Unlike trained lip-synced speech editing, zero-shot speech + speaker edits also change who is speaking. $^\dagger$Voice cloning only.}
    \footnotesize
    \setlength{\tabcolsep}{3.4pt}
    \begin{tabular}{llccc}
    \toprule
    Task & Example Instruction & Pre-train & SFT & Eval. \\
    \midrule
    \multicolumn{5}{l}{\textit{Editing}} \\
    General image editing (I2I) & ``Make her hair blue'' & \cmark & \cmark & \S\ref{sec:img_edit} \\
    General video editing (V2V) & ``Turn the car red'' & \cmark & \cmark & \S\ref{sec:vid_edit} \\
    Speech editing$^\dagger$ (A2A) & ``Change the speech to `See you soon''' & \cmark & \cmark & -- \\
    Complex audio editing (A2A) & ``Remove the music; make the rain louder'' & \xmark & \cmark & \S\ref{sec:audio_edit} \\
    Lip-synced speech editing (AV2AV) & ``Change the speech to `We made it out''' & \cmark & \cmark & \S\ref{section:av-speech} \\
    Masked inpainting (AV2AV) & ``Inpaint this video'' & \xmark & \cmark & -- \\
    \midrule
    \multicolumn{5}{l}{\textit{Generation (no AV input)}} \\
    Text-to-X (T2I, T2V, T2A) & ``A dog barking in the rain'' & \cmark & \cmark & -- \\
    Animation (\{I,A,IA\}2\{V,AV\}) & ``A person talking'' & \cmark & \cmark & -- \\
    \midrule
    \multicolumn{5}{l}{\textit{Zero-shot tasks (all AV2AV)}} \\
    Synced AV speech + speaker editing & ``Change the man to a woman, she says...'' & \xmark & \xmark & \S\ref{sec:zeroshot} \\
    Synced audio editing & ``Add a dragging sound effect'' & \xmark & \xmark & \S\ref{sec:zeroshot} \\
    Unsynced audio editing & ``Remove the traffic ambience'' & \xmark & \xmark & \S\ref{sec:zeroshot} \\
    Synced AV scene editing & ``Add more bikes in the background'' & \xmark & \xmark & \S\ref{sec:zeroshot} \\
    Synced video editing & ``Add a person walking to the footsteps'' & \xmark & \xmark & \S\ref{sec:zeroshot} \\
    Unsynced video editing & ``Change the woman's hair to black'' & \xmark & \xmark & \S\ref{sec:zeroshot} \\
    \bottomrule
    \end{tabular}
    \label{tab:tasks}
\vspace{-0.35cm}
\end{table}

\vspace{-0.2cm}
\section{Training for Cross-Modal Transfer} \label{sec: method}
Built on a shared backbone (Section~\ref{sec:arch}), CrossEdit acquires zero-shot AV scene editing from three data sources: mined lip-synced speech editing pairs (Section~\ref{sec: mining}), synthetic audio editing pairs with complex instructions (Section~\ref{sec:synthesis}), and AV masked reconstruction (Section~\ref{sec: inpainting}).
 
\vspace{-0.15cm}
\subsection{Architecture}\label{sec:arch}
Let $\mathcal{M} = \{\mathcal{I}, \mathcal{A}, \mathcal{V}\}$ denote the image ($\mathcal{I} = \mathbb{R}^{H \times W \times 3}$), audio ($\mathcal{A} = \mathbb{R}^{T_a \times 2}$), and video ($\mathcal{V} = \mathbb{R}^{T_v \times H \times W \times 3}$) spaces. We model omni-modal editing as $p_\theta(\hat{Y} \mid Y, X)$, where $X$ is a natural language instruction, and $Y = (y_1, \ldots, y_n)$ and $\hat{Y} = (\hat{y}_1, \ldots, \hat{y}_k)$ are input and output sequences whose elements each belong to some modality $m \in \mathcal{M}$, with their number and modalities varying across examples. We parameterize $p_\theta$ with flow matching~\citep{lipman2023flow} on latent representations of $\hat{Y}$. Using the linear interpolant $\hat{Y}_t = (1 - t)\hat{Y}_0 + t\epsilon$, where $\hat{Y}_0$ is the clean target, $\epsilon \sim \mathcal{N}(0, \mathbf{I})$, and $t \in [0, 1]$, we train a velocity field $v_\theta$ to regress $\epsilon - \hat{Y}_0$:
\begin{equation}
    \mathcal{L} = \mathbb{E}_{(Y, X, \hat{Y}_0), \epsilon, t} \left[ \left\| v_\theta(\hat{Y}_t, t, Y, X) - (\epsilon - \hat{Y}_0) \right\|^2 \right].
    \label{eq:flow_loss}
\end{equation}
We independently replace $Y$ and $X$ with a null token $\varnothing$ at random during training, enabling classifier-free guidance (CFG)~\citep{ho2021classifierfree} (Appendix~\ref{appendix:cfg}). At inference, we initialize $\hat{Y}_1 \sim \mathcal{N}(0, \mathbf{I})$ and integrate from $t=1$ to $t=0$ with an ODE solver.  

In practice, $p_\theta$ is a DiT \citep{peebles2023scalable} MoE \citep{moe, peng2026completemue} (3B active, 12B total parameters). Pre-trained VAEs \citep{vae, polyak2024moviegen, kumar2023high} encode audio (48kHz stereo $\rightarrow$ 256-dim.\ at 40Hz) and visuals (images and video). The visual VAE supports variable resolutions, mapping e.g.\ 512px to 48-dim.\ 32$\times$32 latents at a fixed framerate (24 FPS $\rightarrow$ 6 FPS). Following \citet{wu2025qwen}, text, images, and video also receive clean semantic embeddings from a pre-trained Qwen3-VL \citep{bai2025qwen3}.

\begin{figure}[t]
    \centering
    \includegraphics[width=0.75\linewidth]{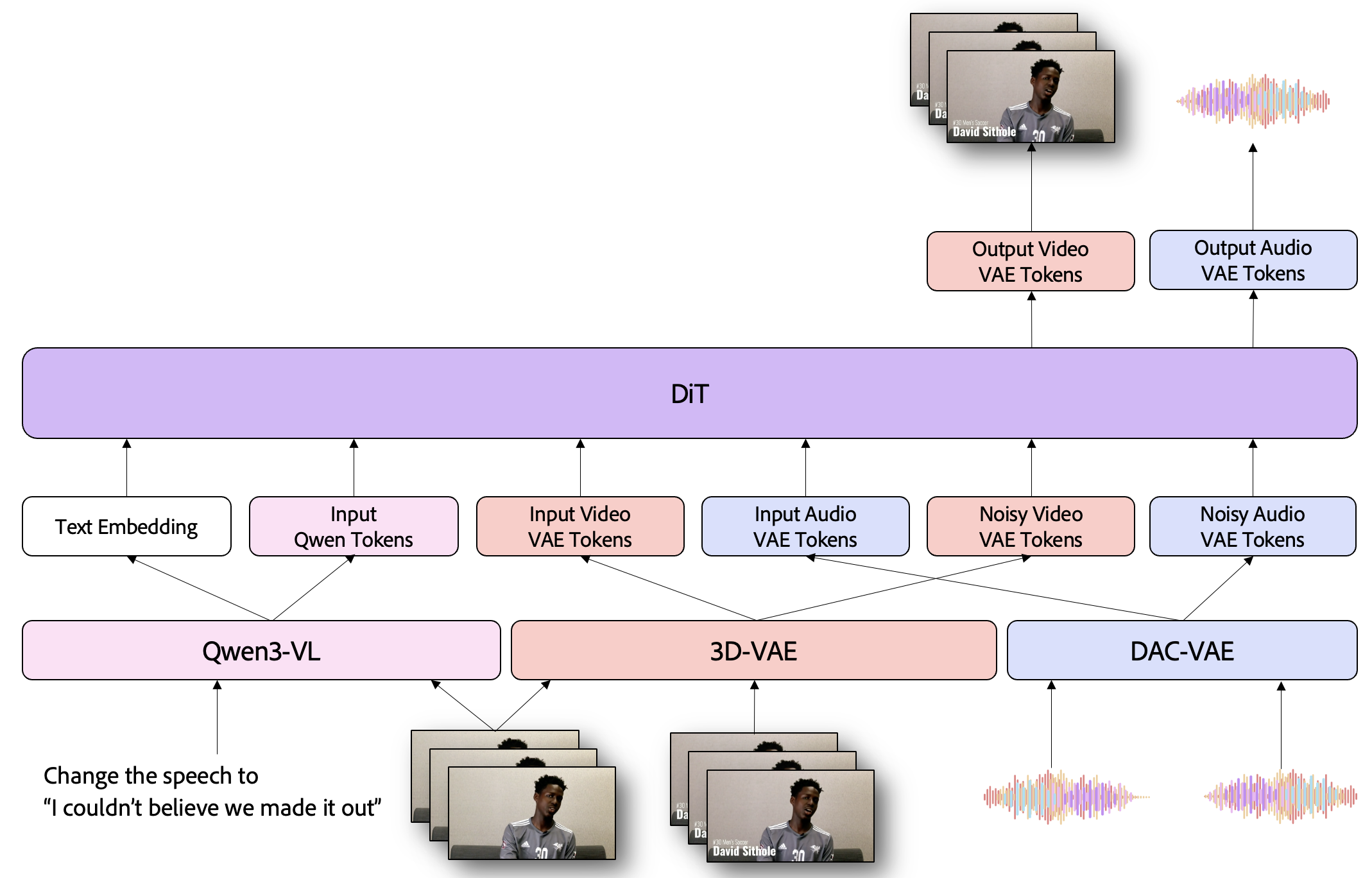}
    \caption{\textbf{Architecture of CrossEdit when encoding AV data.} For simplicity, we omit images, which are treated as a special case of single-frame video.}
    \label{fig:arch}
\vspace{-0.35cm}
\end{figure}
 
\vspace{-0.3cm}
\vspace{-0.15cm}
\subsection{Mining Lip-Synced Speech Editing Pairs} \label{sec: mining}
\vspace{-0.1cm}
 
To teach lip-synced speech editing, we mine pairs in which the same on-screen speaker says different sentences from in-the-wild, commercially usable or licensed videos. These cover only speech content edits, not scene edits (Section~\ref{sec:training_inf}). For each video, we (1) remove non-speech audio with a language-queried source separation model \citep{liu22w_interspeech, wen2026promptsep}; (2) segment and transcribe it with a speaker diarization model similar to \citet{pyannote}, yielding transcripts, speaker labels, and time segments; (3) run face detection and tracking \citep{deng2020retinaface}, discarding segments where a face is missing or goes off-screen; and (4) group retained segments by speaker to form before/after pairs. Filtering discards $\sim$96\% of the source data, but since each speaker yields many pairs, the overall yield is $\sim$20\% (e.g., 100K clips yield 20K pairs).
\vspace{-0.1cm}
\vspace{-0.15cm}
\subsection{Audio Editing Data Synthesis} \label{sec:synthesis}
\vspace{-0.1cm}
Mined data alone is small, biased toward clean speech, and often inconsistent between input and output (e.g., differing background noise). We therefore complement it with synthetic audio editing data from two sources: (1) random mixtures of 2--4 audio files, each from one of four classes (speech, music, background, sound effect), following \citet{tac}, which give acoustically plausible but semantically incoherent scenes of real audio; and (2) scenes synthesized by a tool-calling LLM equipped with generative audio models, following \citet{chen2026audiochat}, which yield more natural audio scenes that are more similar to that found in the real world, but with synthetically generated audio. Combining both sources yields a large-scale dataset of compositional instructions that overcomes the weaknesses of the mined data and helps the model generalize to unseen acoustic scenes; as we show in Section~\ref{sec:zeroshot}, the instruction following it teaches also transfers to AV movie scene editing.

\begin{figure}[t]
    \centering
    \includegraphics[width=0.9\linewidth]{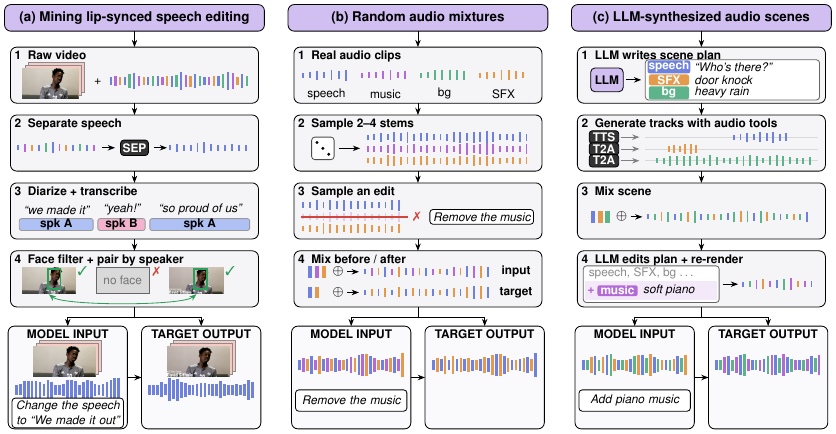}
    \caption{\textbf{Editing data sources.} (a) Our pipeline for mining lip-synced speech editing pairs. (b) Random audio mixtures and (c) Synthetic audio scenes provide audio editing pairs.}
    \label{fig:data}
\vspace{-0.6cm}
\end{figure}
\vspace{-0.1cm}
\vspace{-0.15cm}
\subsection{Audio-Visual Masked Reconstruction} \label{sec: inpainting}
\vspace{-0.1cm}
While mined and synthetic pairs teach instruction following, neither requires the model to generate new visual content within an existing AV scene, and we find them insufficient for complex scene edits (Figure~\ref{fig:demo}). Inspired by video generation \citep{voleti2022mcvd} and cross-modal representation learning \citep{shi2022learning, hsu2022u, gong2022contrastive}, we frame AV masking as a self-supervised AV2AV editing task (Figure~\ref{fig:masking}). Each AV sample receives one of three maskings: \textit{temporal}, masking a random span of frames (and, 50\% of the time, its audio); \textit{spatial}, applying $s \sim \mathcal{U}\{0, S\}$ masks of 200$\times$200px to all frames, with $S{=}12$ for our 540p videos; or \textit{spatio-temporal}, applying both. The model must reconstruct the unmasked clip given the instruction ``inpaint this video''.
 
\vspace{-0.15cm}
\subsection{Training} \label{sec:training_inf}
Given the scarcity of editing data, we start from an internal DiT checkpoint pre-trained on the tasks in Table~\ref{tab:tasks}, which we call \textbf{CrossEdit-Base}\footnote{Pre-training includes the mined lip-synced speech editing pairs (Section~\ref{sec: mining}) but not the synthetic audio data (Section~\ref{sec:synthesis}) or masked reconstruction (Section~\ref{sec: inpainting}).}. CrossEdit-Base is a performant omni-modal editing model on its own, but it cannot consistently perform our target task of AV2AV movie scene editing. We add the synthetic audio editing and masked reconstruction data and fine-tune for 10K steps to obtain \textbf{CrossEdit} (data details, training settings, and inference hyperparameters in Appendix~\ref{appendix:sft}).

Our central goal is to perform AV2AV editing of complex movie scenes without any AV2AV training data for those types of instructions. Table~\ref{tab:tasks} lists every task CrossEdit sees during pre-training (aka CrossEdit-Base) and supervised fine-tuning (SFT). In all of its training, CrossEdit sees only two AV2AV instruction types: lip-synced speech editing (Section~\ref{sec: mining}) and masked inpainting (Section~\ref{sec: inpainting}). The instructions in Figure~\ref{fig:demo} (adding objects, removing ambience, changing speaker attributes) appear only in single-modality editing data (I2I, V2V, A2A), so AV2AV editing with them is \textit{zero-shot}: it requires transferring instruction following learned in other modalities. Section~\ref{sec:zeroshot} evaluates these zero-shot AV edits, and Section~\ref{sec:trained} checks that our techniques maintain or improve trained-task performance. 

\vspace{-0.2cm}
\section{Zero-Shot Audio-Visual Scene Editing} \label{sec:zeroshot}
We now evaluate our central claim that cross-modal training enables AV movie scene editing without AV scene editing data, using our proposed benchmark (Section~\ref{sec:benchmark}) and metric (Section~\ref{sec:avfes}).
 
\vspace{-0.15cm}
\subsection{CrossEditBench} \label{sec:benchmark}
As no public benchmark exists for instruction-based AV movie scene editing, we release \textbf{CrossEditBench}: 100 clips from the evaluation split of OSSL \citep{ossl}, which is built from public domain movies, each annotated by humans with an edit instruction. Annotators also specify what should change in each modality (e.g., ``add a cow and remove the music'' $\rightarrow$ audio: [add cow mooing, remove music]; video: [add cow]), which we release for future fine-grained evaluation.
 
Instructions vary along two axes (Figure~\ref{fig:demo}): \textit{which modality the edit targets} and \textit{whether it must be synchronized across modalities}. An \textit{unsynced} edit must leave the other modality untouched (removing traffic ambience must not alter the video), while a \textit{synced} edit must align with it (an added dragging sound must start and stop with the on-screen motion). This yields six categories (Table~\ref{tab:tasks}): unsynced audio, synced audio, unsynced video, synced video, synced AV scene, and synced AV speech and speaker editing. As none of them appear in AV2AV training data (Section~\ref{sec:training_inf}), CrossEditBench directly measures zero-shot transfer. More examples are in Appendix~\ref{appendix:benchmark}.

\begin{figure}
    \centering
    \includegraphics[width=0.4\linewidth]{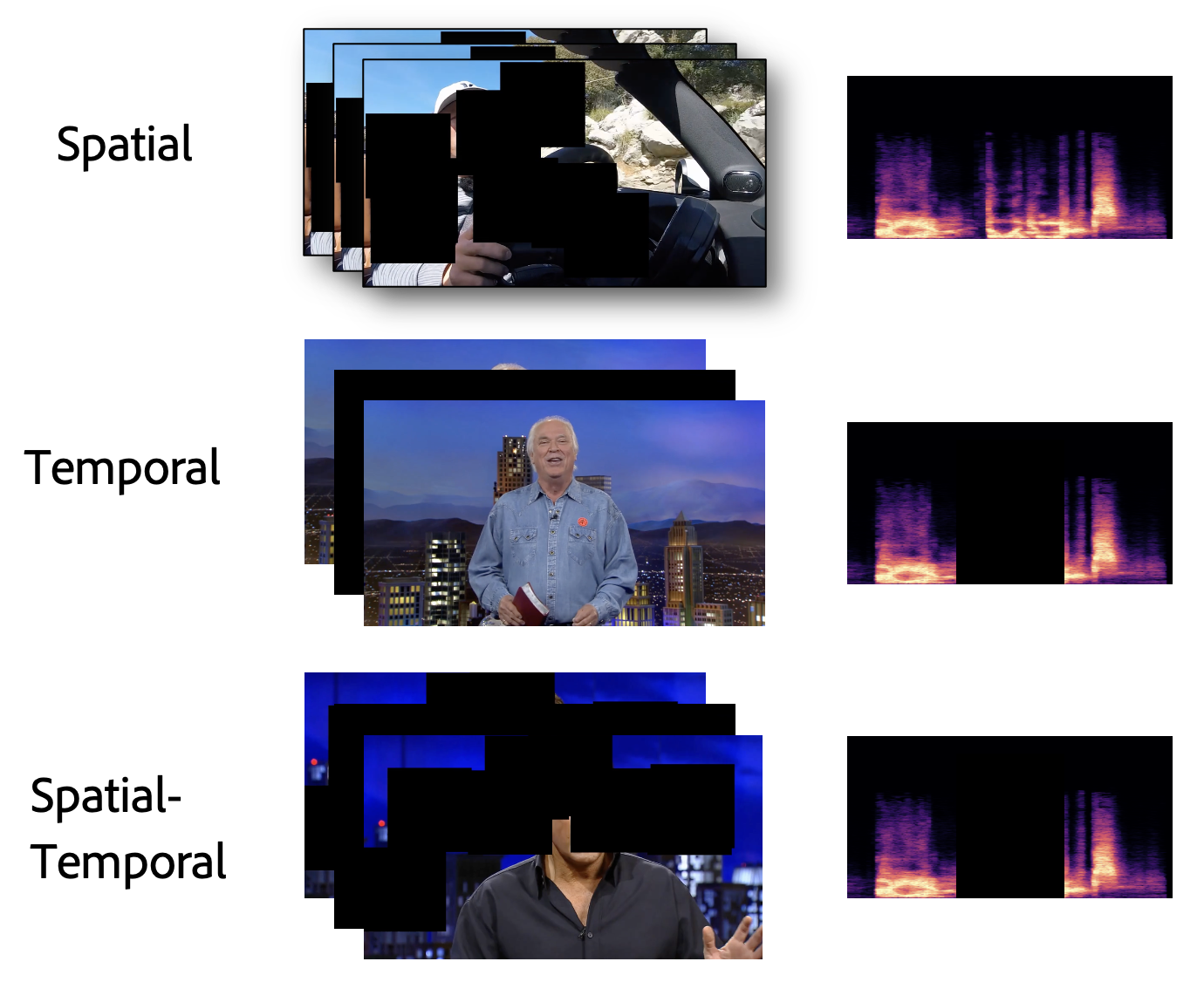}
    \caption{\textbf{AV masked reconstruction schemes.} All use the instruction ``Inpaint this video''.}
    \label{fig:masking}
    \vspace{-0.45cm}
\end{figure}
 
\vspace{-0.15cm}
\subsection{AV-FES: Audio-Visual Faithful Edit Score} \label{sec:avfes}
Existing automatic metrics for AV editing score the edited output in isolation from the input or instruction and often remain disjoint across modalities \citep{Girdhar_2023_CVPR, aved, zheng2025avi, liang2026spongebob}.  Editing, as a highly conditional task, requires comparing the output against both the instruction and the input. AV-FES is grounded in two fundamental sub-scores often used in uni-modal editing evaluations, Instruction Following and Consistency \citep{chen2026audiochat, guo2025audiostory, ma2026mmae}, with respect to the text instruction and input media. While both are often evaluated on a 5-point Likert scale by a judge, such as a human annotator or LLM, the two axes pull in opposite directions. Leaving the input untouched maximizes consistency, while re-synthesizing the whole clip is the easiest way to satisfy the instruction. This makes it difficult to evaluate editing systems: how does one weigh gains in one sub-score against losses in the other?
 
We observed that this relationship is similar to the sub-scores of the commonly used $F_1$ metric: Precision (akin to Consistency) and Recall (parallel to Instruction Following). Motivated by this analogy, we therefore combine the two judge axes into a single score, the Audio-Visual Faithful Edit Score (AV-FES), computed separately for the video (V-FES) and audio (A-FES) modalities and averaged. For each input clip $Y$ with instruction $X$, an LLM judge rates a system output $\hat{Y}$ on instruction following with respect to $X$, $\mathrm{Ins}_m(\hat{Y}) \in \{1,\dots,5\}$, and consistency with respect to $Y$, $\mathrm{Cons}_m(\hat{Y}) \in \{1,\dots,5\}$, for each modality $m \in \{\mathcal{V}, \mathcal{A}\}$. We also rate the unedited input $Y$ itself under the same instruction, yielding $\mathrm{Ins}_m(Y)$: the score a system obtains by doing nothing.
 
\paragraph{Instruction Gain.}
We first calculate the gain $g_m(\hat{Y})$, which quantifies the amount of achievable improvement the system obtained over the unedited input in Instruction Following:
\begin{equation}
    g_m(\hat{Y}) \;=\; \operatorname{clip}_{[0,1]}\!\left(
        \frac{\mathrm{Ins}_m(\hat{Y}) - \mathrm{Ins}_m(Y)}{5 - \mathrm{Ins}_m(Y)}
    \right),
    \label{eq:gain}
\end{equation}
so $g_m = 0$ for any output the judge rates no better than the input (copying), and $g_m = 1$
when the output reaches the maximum score. Normalizing by the per-clip headroom
$5 - \mathrm{Ins}_m(Y)$ makes clips comparable regardless of how much the judge credits the
untouched input.

\paragraph{Consistency.}
To make the sub-scores comparable, we then map the Consistency rating to the unit interval, $c_m(\hat{Y}) \;=\; \frac{\mathrm{Cons}_m(\hat{Y}) - 1}{4}$, so that ratings of 5 and 1 map to 1 and 0.
 
\paragraph{Faithful Edit Score.}
The per-modality FES is the harmonic mean of gain and consistency:
\begin{equation}
    \text{FES}_m(\hat{Y}) \;=\;
    \frac{2\, g_m(\hat{Y})\, c_m(\hat{Y})}{g_m(\hat{Y}) + c_m(\hat{Y})},
    \label{eq:fes}
\end{equation}
with $\text{FES}_m(\hat{Y}) = 0$ when $g_m(\hat{Y}) = c_m(\hat{Y}) = 0$.
A straight-through copy of the input would score near $0$ since $g_m \approx 0$, whereas a full re-synthesis that happens to satisfy the instruction is dragged down by a low $c_m$. With V-FES $= \text{FES}_\mathcal{V}$ and A-FES $= \text{FES}_\mathcal{A}$, the overall score for AV editing is their average:
\begin{equation}
    \text{AV-FES}(\hat{Y}) \;=\; \tfrac{1}{2}\left(\text{FES}_\mathcal{V}(\hat{Y}) + \text{FES}_\mathcal{A}(\hat{Y})\right).
    \label{eq:avfes}
\end{equation}
All scores are computed per clip. AV-FES therefore rewards edits for each modality individually while penalizing unintended changes within a single metric. Additional details are in Appendix~\ref{appendix:avfes}.
 
\vspace{-0.15cm}
\subsection{Results} \label{sec:zeroshot_results}
\textbf{Setup:} We evaluate CrossEdit's ability to edit entire AV scenes in a zero-shot manner (Section~\ref{sec:training_inf}) using CrossEditBench (Section \ref{sec:benchmark}). We report AV-FES using Gemini 3.1 Pro (Section~\ref{sec:avfes}) and AV alignment via ImageBind score \citep{Girdhar_2023_CVPR}. We compare to two concurrent end-to-end models, (InstructAV2AV \citep{zheng2026instructav2av} and the production-grade, 33B-parameter MiniMax H3 \citep{minimax2026h3}), modular cascades (Lucy Edit \citep{lucyedit} (V2V) + HY-Video-Foley XL/XXL \citep{shan2025hunyuanvideofoleymultimodaldiffusionrepresentation} (V2AV) and VACE (V2V) + Coherent (VA2A) \citep{jiang2025vace, ishii2025coherent}), an agentic system (AVI-EDIT \citep{zheng2025avi}), and zero-shot delta denoising (AvED \citep{aved}). Subjective test details are in Appendix~\ref{appendix: subjective}.
 
\textbf{Results:} CrossEdit achieves the highest AV-FES, V-FES, and A-FES, driven by the strongest instruction following in both modalities (Table~\ref{tab:scene_edit}). It outperforms both concurrent end-to-end models, including MiniMax H3 (AV-FES 0.30 vs.\ 0.23) despite being under half its size (12B vs.\ 33B parameters), which is competitive on video (V-FES 0.24 vs.\ 0.26) but under-edits the audio (audio instruction following 2.77 vs.\ 3.70); InstructAV2AV trails on both modalities (AV-FES 0.20). Cascades lack cross-modal context, so the audio stage re-synthesizes the soundtrack and audio consistency collapses. AvED stays close to the input, attaining the highest ImageBind score but among the lowest gains, while AVI-EDIT degrades the video. CrossEdit also improves over CrossEdit-Base on every FES score (Section~\ref{sec:ablation}). Finally, our conclusions are robust to the choice of judge: Qwen3-Omni \citep{xu2025qwen3} correlates with Gemini at 0.90 (Pearson and Spearman) on system-level AV-FES, and both judges' rankings closely match human ratings (AV-FES Pearson 0.92 and 0.96), albeit over only 4--5 shared systems (Appendix~\ref{appendix:judge}). Example outputs are in Appendix~\ref{appendix:benchmark}.
 
\begin{table}[t]
    \centering
    \caption{\textbf{CrossEditBench AV zero-shot scene editing results.} \textcolor{gray}{Gray} columns are the modality-specific FES subscores. Improvements due to our proposed techniques are \underline{underlined}.}
    \resizebox{\textwidth}{!}{
    \begin{tabular}{lc|c|c ccc|c ccc|c}
    \toprule
     & & \textbf{Overall} & \multicolumn{4}{c|}{Visual} & \multicolumn{4}{c|}{Audio} & Alignment \\
     & & & & \multicolumn{3}{c|}{\textcolor{gray}{\textit{subscores}}} & & \multicolumn{3}{c|}{\textcolor{gray}{\textit{subscores}}} & \\
     \cmidrule(lr){5-7} \cmidrule(lr){9-11}
     & Type & \textbf{AV-FES} $\uparrow$ & \textbf{V-FES} $\uparrow$ & \textcolor{gray}{Gain} & \textcolor{gray}{Ins.} & \textcolor{gray}{Cons.} & \textbf{A-FES} $\uparrow$ & \textcolor{gray}{Gain} & \textcolor{gray}{Ins.} & \textcolor{gray}{Cons.} & ImageBind $\uparrow$ \\
    \midrule
    Input Video & & -- & -- & \textcolor{gray}{--} & \textcolor{gray}{2.81} & \textcolor{gray}{3.97} & -- & \textcolor{gray}{--} & \textcolor{gray}{2.19} & \textcolor{gray}{3.95} & 19.79 \\
    \midrule
    LucyEdit + HY-XL  & cascade  & 0.16 & 0.12 & \textcolor{gray}{0.12} & \textcolor{gray}{2.88} & \textcolor{gray}{3.76} & 0.19 & \textcolor{gray}{0.25} & \textcolor{gray}{2.79} & \textcolor{gray}{2.27} &  9.81 \\
    LucyEdit + HY-XXL & cascade  & 0.16 & 0.12 & \textcolor{gray}{0.12} & \textcolor{gray}{2.88} & \textcolor{gray}{3.76} & 0.20 & \textcolor{gray}{0.29} & \textcolor{gray}{2.89} & \textcolor{gray}{2.15} & 13.20 \\
    VACE + Coherent   & cascade  & 0.19 & 0.12 & \textcolor{gray}{0.14} & \textcolor{gray}{3.16} & \textcolor{gray}{3.44} & 0.26 & \textcolor{gray}{0.34} & \textcolor{gray}{3.12} & \textcolor{gray}{2.49} & 13.22 \\
    AvED              & sampling & 0.12 & 0.09 & \textcolor{gray}{0.10} & \textcolor{gray}{2.81} & \textcolor{gray}{3.20} & 0.14 & \textcolor{gray}{0.13} & \textcolor{gray}{2.49} & \textcolor{gray}{3.93} & \textbf{19.38} \\
    AVI-EDIT          & agentic  & 0.17 & 0.08 & \textcolor{gray}{0.10} & \textcolor{gray}{2.42} & \textcolor{gray}{2.35} & 0.25 & \textcolor{gray}{0.27} & \textcolor{gray}{2.93} & \textcolor{gray}{3.65} & 15.45 \\
    \midrule
    InstructAV2AV     & e2e      & 0.20 & 0.15 & \textcolor{gray}{0.15} & \textcolor{gray}{2.85} & \textcolor{gray}{3.03} & 0.24 & \textcolor{gray}{0.26} & \textcolor{gray}{2.85} & \textcolor{gray}{3.20} & 16.08 \\
    MiniMax H3        & e2e      & 0.23 & 0.24 & \textcolor{gray}{0.23} & \textcolor{gray}{3.14} & \textcolor{gray}{3.38} & 0.22 & \textcolor{gray}{0.22} & \textcolor{gray}{2.77} & \textcolor{gray}{3.70} & 18.90 \\
    \midrule
    CrossEdit-Base    & e2e      & 0.26 & 0.20 & \textcolor{gray}{0.24} & \textcolor{gray}{2.88} & \textcolor{gray}{2.00} & 0.31 & \textcolor{gray}{0.39} & \textcolor{gray}{3.15} & \textcolor{gray}{3.52} & 14.30 \\
    CrossEdit         & e2e      & \underline{\textbf{0.30}} & \underline{\textbf{0.26}} & \textcolor{gray}{0.28} & \textcolor{gray}{3.59} & \textcolor{gray}{2.81} & \underline{\textbf{0.33}} & \textcolor{gray}{0.35} & \textcolor{gray}{3.70} & \textcolor{gray}{3.84} & \underline{16.29} \\
    \bottomrule
    \end{tabular}}
    \label{tab:scene_edit}
\vspace{-0.35cm}
\end{table}
 
\vspace{-0.15cm}
\subsection{Ablation of Training Methods} \label{sec:ablation}
Table~\ref{tab:data_ablation} reports human ratings, and the FES computed from them, as each training component is added. CrossEdit-Base attempts edits but often alters unrelated content, giving the lowest visual consistency of all systems. Adding synthetic audio editing data, which contains no visual content, improves instruction following in both modalities as well as visual consistency, raising AV-FES from 0.26 to 0.35, consistent with cross-modal transfer. AV masked reconstruction further improves all four axes, reaching 0.43. The cascades best preserve the video but barely follow visual instructions and overwrite the original soundtrack (AV-FES 0.11--0.14). On lip-synced speech editing, audio editing data likewise improves visual consistency (Appendix~\ref{appendix: subjective}).

\begin{figure}[t]
    \centering
    \includegraphics[width=0.8\linewidth]{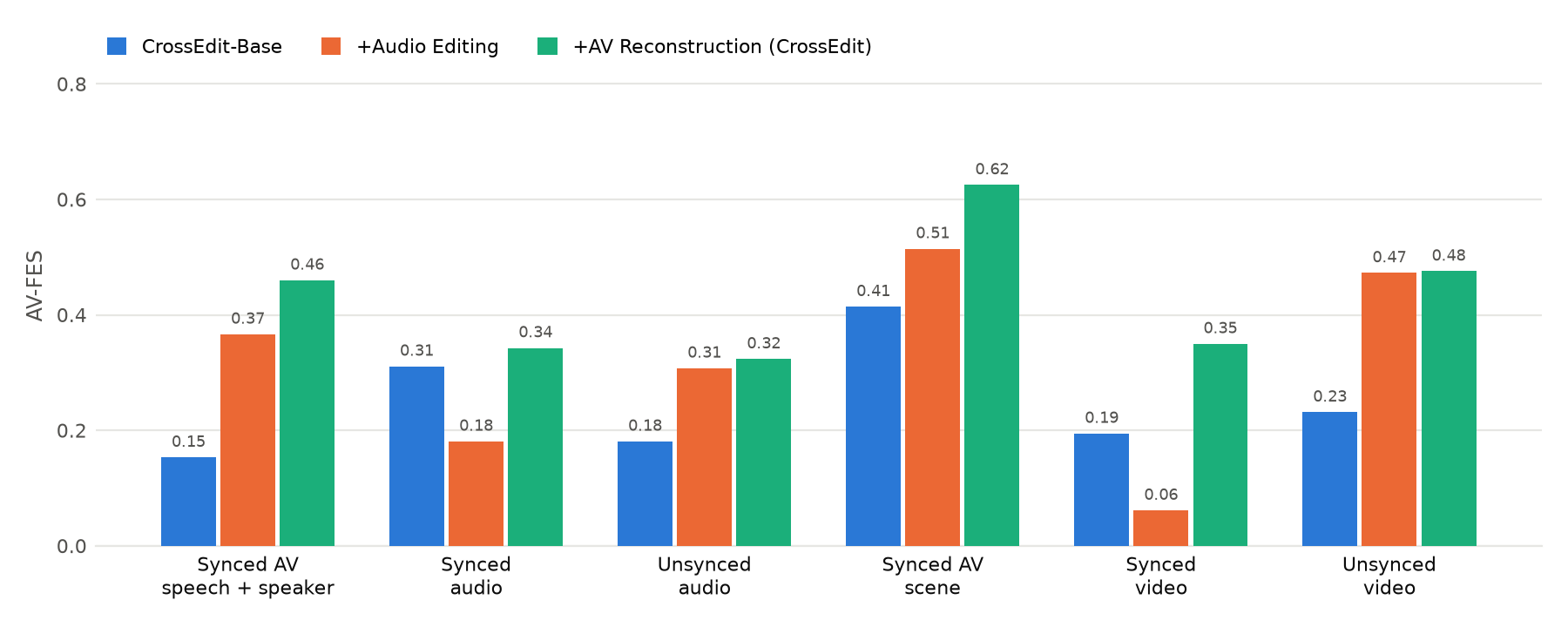}
    \caption{\textbf{Human-rated AV-FES on CrossEditBench, by instruction category.}}
    \label{fig:category}
\vspace{-0.35cm}
\end{figure}

Figure~\ref{fig:category} breaks down human-rated AV-FES by type. CrossEdit improves over CrossEdit-Base in every category, with the largest gains on synced AV scene edits (0.41 to 0.62), speech+speaker edits (0.15 to 0.46), and unsynced video edits (0.23 to 0.48). Audio editing data alone helps most on edits that mainly require following the instruction, but hurts synced audio and synced video edits (0.31 to 0.18 and 0.19 to 0.06), where the new content must be generated in time with the other modality. Adding AV masked reconstruction recovers and surpasses CrossEdit-Base on both (0.34 and 0.35), consistent with its role of teaching the model to generate content that fits the existing scene.
 
\begin{table}[t]
    \centering
    \caption{\textbf{Ablation on zero-shot AV scene editing with human ratings} (344 per system; Ins.\ and Cons.\ with 95\% CIs). Improvements over CrossEdit-Base are \underline{underlined}.}
    \resizebox{\textwidth}{!}{
    \begin{tabular}{l|c|c ccc|c ccc}
    \toprule
     & \textbf{Overall} & \multicolumn{4}{c|}{Visual} & \multicolumn{4}{c}{Audio} \\
     & & & \multicolumn{3}{c|}{\textcolor{gray}{\textit{subscores}}} & & \multicolumn{3}{c}{\textcolor{gray}{\textit{subscores}}} \\
     \cmidrule(lr){4-6} \cmidrule(lr){8-10}
     & \textbf{AV-FES} $\uparrow$ & \textbf{V-FES} $\uparrow$ & \textcolor{gray}{Gain} & \textcolor{gray}{Ins.} & \textcolor{gray}{Cons.} & \textbf{A-FES} $\uparrow$ & \textcolor{gray}{Gain} & \textcolor{gray}{Ins.} & \textcolor{gray}{Cons.} \\
    \midrule
    Input & -- & -- & \textcolor{gray}{--} & \textcolor{gray}{2.75$\pm$0.24} & \textcolor{gray}{4.71$\pm$0.11} & -- & \textcolor{gray}{--} & \textcolor{gray}{2.29$\pm$0.25} & \textcolor{gray}{4.68$\pm$0.11} \\
    \midrule
    LucyEdit + HY-XL & 0.11 & 0.14 & \textcolor{gray}{0.12} & \textcolor{gray}{2.61$\pm$0.25} & \textcolor{gray}{3.93$\pm$0.20} & 0.08 & \textcolor{gray}{0.16} & \textcolor{gray}{2.17$\pm$0.20} & \textcolor{gray}{1.74$\pm$0.20} \\
    LucyEdit + HY-XXL & 0.14 & 0.17 & \textcolor{gray}{0.15} & \textcolor{gray}{2.74$\pm$0.21} & \textcolor{gray}{3.58$\pm$0.20} & 0.12 & \textcolor{gray}{0.17} & \textcolor{gray}{2.22$\pm$0.20} & \textcolor{gray}{1.95$\pm$0.21} \\
    \midrule
    CrossEdit-Base & 0.26 & 0.25 & \textcolor{gray}{0.34} & \textcolor{gray}{2.85$\pm$0.20} & \textcolor{gray}{2.37$\pm$0.19} & 0.27 & \textcolor{gray}{0.32} & \textcolor{gray}{2.89$\pm$0.21} & \textcolor{gray}{2.80$\pm$0.21} \\
    \hspace{0.3cm} + Audio Editing & \underline{0.35} & \underline{0.35} & \textcolor{gray}{0.43} & \textcolor{gray}{3.31$\pm$0.21} & \textcolor{gray}{2.94$\pm$0.18} & \underline{0.35} & \textcolor{gray}{0.46} & \textcolor{gray}{3.25$\pm$0.21} & \textcolor{gray}{2.81$\pm$0.19} \\
    \hspace{0.7cm}+ AV Masked Reconstruction (CrossEdit) & \underline{\textbf{0.43}} & \underline{\textbf{0.42}} & \textcolor{gray}{0.47} & \textcolor{gray}{3.48$\pm$0.18} & \textcolor{gray}{3.21$\pm$0.18} & \underline{\textbf{0.44}} & \textcolor{gray}{0.47} & \textcolor{gray}{3.44$\pm$0.18} & \textcolor{gray}{3.44$\pm$0.20} \\
    \bottomrule
    \end{tabular}}
    \label{tab:data_ablation}
\vspace{-0.35cm}
\end{table}

\vspace{-0.2cm}
\section{Retaining Omni-Modal Performance} \label{sec:trained}
Having shown that cross-modal training enables zero-shot AV scene editing, we verify that it does not come at the expense of the lip-synced speech editing, image editing, and video editing tasks CrossEdit was trained on. We note that we primarily focus on the comparison betwee CrossEdit and\ CrossEdit-Base, with other models shown for reference. Results on audio editing are in Appendix~\ref{sec:audio_edit}.
 
\vspace{-0.15cm}
\subsection{Lip-Synced Speech Editing} \label{section:av-speech}
\textbf{Setup:} We use an internal set of 600 IMDB celebrity interviews. Given a new spoken sentence, the model must regenerate the audio and video with correspondingly synchronized lip movements and facial expressions (Figure~\ref{fig:arch}). We measure WER and WavLM \citep{ChenWavLm} speaker similarity (SPK) to the input; DINO \citep{oquab2024dinov} similarity to the input and across frames; AV sync via LSE-D and LSE-C \citep{syncnet}; and human ratings (Appendix~\ref{appendix: subjective}). As no public end-to-end system exists, we compare to cascades of F5-TTS \citep{chen-etal-2025-f5} with HuMo \citep{chen2025humo} or StableAvatar \citep{tu2025stableavatar}, which animate the first input frame.
 
\textbf{Results:} Our SFT improves speaker similarity, video input consistency, and LSE-D, with minor gains in temporal consistency (Table~\ref{tab:av_speech_editing}). F5-TTS remains stronger on WER and SPK as a dedicated TTS system, but the cascades cannot preserve the input video, and human raters strongly prefer both CrossEdit models, especially for video consistency. Qualitatively, SFT mainly improves preservation of the input's visual background (Figure~\ref{fig:consistency} in Appendix~\ref{appendix:speech_edit}), reflected in higher video consistency. Our ablation (Table~\ref{tab:data_ablation_appendix_speech}) attributes most of this gain to the audio editing data: we hypothesize that simulated audio editing pairs teach the model to edit only what is necessary.
 
\begin{table}[t]
    \centering
    \caption{\textbf{Lip-synced speech editing results.} Improvements due to our techniques are \underline{underlined}.}
    \resizebox{\textwidth}{!}{
    \begin{tabular}{l|cc|cc|cc|cccc}
    \toprule
     & \multicolumn{2}{c|}{Speech} & \multicolumn{2}{c|}{Video} & \multicolumn{2}{c|}{Sync} & \multicolumn{4}{c}{Human Ratings} \\
     & & & & & & & \multicolumn{2}{c}{Audio} & \multicolumn{2}{c}{Video} \\
     & WER $\downarrow$ & SPK $\uparrow$ & Time $\uparrow$ & Input $\uparrow$ & LSE-D $\downarrow$ & LSE-C $\uparrow$ & Ins. $\uparrow$ & Cons. $\uparrow$ & Ins. $\uparrow$ & Cons. $\uparrow$ \\
     \midrule
     Input & - & - & - & - & - & - & 3.29 & 4.39 & 3.29 & 4.50 \\
     \midrule
     F5-TTS & \textbf{7.5} & \textbf{89.3} & - & - & - & - & - & - & - & - \\
     + HuMo & - & - &  99.1 & 67.4 & 9.12 & \textbf{5.88} & 3.30 & 2.97 & 2.60 & 2.10 \\
     + StableAvatar & - & - & 97.5 & 42.0 & 12.45 & 2.13 & 3.21 & 2.88 & 2.17 & 1.78 \\
     \midrule
     CrossEdit-Base & 10.3 & 86.9 & 99.1 & 90.0 & 9.00 & 5.87 & \textbf{3.98} & 3.81 & \textbf{4.05} & 4.22 \\
     CrossEdit & 10.9 & \underline{88.5} & \underline{\textbf{99.2}} & \underline{\textbf{92.1}} & \underline{\textbf{8.92}} & 5.64 & 3.96 & \underline{\textbf{3.82}} & 4.02 & \underline{\textbf{4.42}} \\
     \bottomrule
    \end{tabular}}
    \label{tab:av_speech_editing}
\vspace{-0.35cm}
\end{table}

\vspace{-0.15cm}
\subsection{Video Editing} \label{sec:vid_edit}
\textbf{Setup:} We evaluate on EditVerseBench \citep{ju2026editverse} with GPT-4o-judged editing quality \citep{hurst2024gpt}, PickScore, text alignment, and CLIP/DINO temporal consistency. Our goal is to retain CrossEdit-Base's performance: dedicated video editing models (InsV2V \citep{cheng2024consistent}, Lucy Edit \citep{lucyedit}, and EditVerse \citep{ju2026editverse}) are included as reference points.
 
\textbf{Results:} Rather than merely retaining video editing performance, our training improves 5 of 6 metrics over CrossEdit-Base (Table~\ref{tab:editverse}), and CrossEdit remains in the range of dedicated models such as Lucy Edit. We hypothesize that the synthetic audio data exposes the model to more diverse instructions.
 
\begin{table}[th]
    \centering
    \caption{\textbf{Video editing on EditVerseBench.} Improvements from our techniques are \underline{underlined}.}
    \begin{tabular}{l|c|c|cc|cc}
    \toprule
        & \multicolumn{1}{c}{VLM Judge} & \multicolumn{1}{c}{Quality} & \multicolumn{2}{c}{Text Align.} & \multicolumn{2}{c}{Temporal Cons.} \\
         & Editing Quality $\uparrow$ & Pick Score $\uparrow$  & Frame $\uparrow$  & Video $\uparrow$  & CLIP $\uparrow$  & DINO $\uparrow$ \\
    \midrule
     InsV2V & 5.21	& 19.39	& 24.99	& 22.54	& 97.15	& 96.57\\
     Lucy Edit & 5.89 & 19.67 &  26.00 &  23.11 &  98.49 &  98.38\\
     EditVerse & \textbf{7.65} & \textbf{20.07} & \textbf{26.73} & \textbf{23.93} & \textbf{98.56} & \textbf{98.40}  \\
     \midrule
     CrossEdit-Base & 6.86 & 19.42 & 23.10 & 20.10 & 98.00 & 96.90 \\
     CrossEdit & 6.85 & \underline{19.82} & \underline{25.21} & \underline{22.26} & \underline{98.43} & \underline{97.52}\\
    \bottomrule
    \end{tabular}
    \label{tab:editverse}
\vspace{-0.35cm}
\end{table}
 
\vspace{-0.15cm}
\subsection{Image Editing} \label{sec:img_edit}
\textbf{Setup:} We use ImgEdit \citep{ye2026imgedit}, which scores 9 editing categories with a GPT-4o judge. As with video, our goal is to retain CrossEdit-Base's performance; open SOTA image editing models \citep{wu2025omnigen2,wu2025qwen, lin2025uniworld,bagel} are included only as reference points.
 
\textbf{Results:} Adding audio data and AV masked reconstruction slightly improves the overall score over CrossEdit-Base (3.69 to 3.72), with large gains in the Action, Adjust, Hybrid, and Remove categories (Table~\ref{tab:img_edit}). Both remain competitive with dedicated image models of similar or larger size.

\begin{table}[th]
    \centering
    \caption{\textbf{Image editing results on ImgEdit (1--5, higher is better)}. Dedicated image editing models are shown for reference, with scores from \citet{wu2025qwen}}
    \resizebox{\textwidth}{!}{
    \begin{tabular}{lc|ccccccccc|c}
    \toprule
         & Param. & Add & Adjust & Extract & Replace & Remove &  Background & Style & Hybrid & Action & Overall \\
     \midrule
     BAGEL & 14B & 3.56 & 3.31 & 1.70 & 3.30 & 2.62 & 3.24 &  4.49 & 2.38 & 4.17 & 3.20 \\
     UniWorld-V1 & 12B & 3.82 & 3.64 & 2.27 & 3.47 & 3.24 &  2.99 & 4.21 & 2.96 & 2.74 & 3.26  \\
     OmniGen2 & 7B & 3.57 & 3.06 & 1.77 & 3.74 & 3.20 & 3.57 & 4.81 & 2.52 & 4.68 & 3.44 \\
     Qwen-Image & 20B  &  4.38 & 4.16 & \textbf{3.43} & \textbf{4.66} & 4.14 & \textbf{4.38} & 4.81 & 3.82 & 4.69  & \textbf{4.27}\\
     \midrule
     CrossEdit-Base & 12B & 4.37 & 4.06 & 1.48 & 3.61 & 3.43 & 3.88 & 4.54 & 3.22 & 4.48 & 3.69\\
     CrossEdit & 12B & 4.20 & \underline{4.29} & \underline{1.74} & 3.61 & \underline{\textbf{4.32}} & 2.82 & 4.44 & \underline{3.57} & \underline{\textbf{4.96}} & \underline{3.72}\\
     \bottomrule
    \end{tabular}}
    \label{tab:img_edit}
\vspace{-0.35cm}
\end{table}

\vspace{-0.2cm}
\section{Conclusion}
We presented CrossEdit, an omni-modal editing model that edits real movie scenes from free-form instructions zero-shot, enabled by two complementary techniques: cross-modal transfer of instruction following from synthetic audio editing data, and AV masked reconstruction as a self-supervised editing task. We also introduced CrossEditBench and the AV-FES metric for evaluating instruction-based AV movie scene editing. CrossEdit outperforms cascaded, sampling-based, agentic, and SOTA end-to-end models on these tasks while maintaining or improving performance on lip-synced speech, image, and video editing.

\clearpage

\subsection*{AI use statement}

In this work, we used generative AI tools to generate synthetic datasets, refine hypotheses, design or provide feedback on research  methodology or experiments, implement methods, support qualitative and thematic data analysis, and to interpret results. 

We have not used generative AI tools to help develop theoretical models or conceptual frameworks, formulate mathematical claims, provide critical ingredients for proving mathematical claims, assist in the writing of proofs, assist with translation, or clean and reformat dataset. 

Additionally, we used generative AI tools to create or modify scientific figures or images, create or edit software code, create artifacts, draft parts of a research paper, search for information, edit a research paper to improve readability, and identify relevant literature.

We have reviewed all AI-assisted work. LLM-generated code, figures, and writing were verified and tested (if applicable) for correctness. We take responsibility for the final content of this work,
including text, claims or artifacts produced with the aid of generative AI.

\subsection*{Ethics statement}

This research can help reduce the barrier to entry for creative uses of multimedia, as professional video and audio editing software can be complex to use. It can also reduce the cost of media production, as editing can be used in place of reshoots. On the other hand, such technologies can affect the labor market or even be used for malicious causes. For example, the lip-synced speech editing capabilities of CrossEdit can often result in outputs that are indistinguishable from real videos, which can then be used for fraud, impersonation, or libel. We condemn any use of our research for these causes. Due to the risks of this technology, we do not plan on open-sourcing any code or pre-trained checkpoints.

While most of our evaluations are conducted on data that reflects real-world use cases of multi-media editing models (interviews, movie clips, etc.), it does not consider all possible scenarios. For example, we did not test our models on noisy home-made videos with reverberant audio, as those are difficult to obtain annotations for and score. We also only trained and tested the model on short video clips up to 10 seconds long. While it could theoretically work since it was trained on audio up to 30 seconds long, long-form AV editing remains an open challenging task.

\subsection*{Reproducibility statement}

Due to the risks of this technology for making realistic deepfakes and the proprietary nature of the code and training data, the model checkpoints will not be publicly released in their current form. To support reproduction, we describe every component needed to re-implement and evaluate CrossEdit:
\begin{itemize}
    \item \textbf{Model:} the formulation, flow-matching objective, and architecture (Section~\ref{sec:arch}, Appendix~\ref{appendix:architecture}), and classifier-free guidance and sampling hyperparameters (Appendix~\ref{appendix:cfg}).
    \item \textbf{Training data:} the lip-synced speech editing mining pipeline and its yield (Section~\ref{sec: mining}), audio editing data synthesis (Section~\ref{sec:synthesis}), and AV masked reconstruction (Section~\ref{sec: inpainting}, Appendix~\ref{appendix:masking}).
    \item \textbf{Training:} pre-training tasks, SFT data mixture, optimizer, learning rate, and compute (Table~\ref{tab:tasks}, Appendix~\ref{appendix:sft}, Table~\ref{tab:sft_data_appendix}).
    \item \textbf{Evaluation:} CrossEditBench construction (Section~\ref{sec:benchmark}, Appendix~\ref{appendix:benchmark}); the AV-FES definition, including edge cases (Section~\ref{sec:avfes}, Appendix~\ref{appendix:avfes}); judge models, rubric, and agreement across judges (Figure~\ref{fig:subjective}, Appendix~\ref{appendix:judge}); the human evaluation protocol (Appendix~\ref{appendix: subjective}); and the lip-synced speech editing and audio editing test sets and metrics (Section~\ref{section:av-speech}, Appendices~\ref{appendix:speech_edit} and~\ref{sec:audio_edit}).
\end{itemize}
We will release the annotated CrossEditBench evaluation set, annotator instructions, judge prompts, and model outputs on the demo page.

\bibliography{iclr2027_conference}
\bibliographystyle{iclr2027_conference}

\appendix

\section{Architecture} \label{appendix:architecture}
CrossEdit consists of frozen modality-specific tokenizers and a latent diffusion backbone \citep{rombach2022high} (Figure~\ref{fig:arch}). Each modality is tokenized by a separate pre-trained VAE \citep{vae}:
\begin{itemize}
    \item \textbf{Audio:} a DAC-VAE \citep{polyak2024moviegen, kumar2023high, adobevae} that encodes 48kHz stereo audio into 256-dimensional continuous latents at 40Hz, pre-trained on over 100K hours of speech, sound effects, and music.
    \item \textbf{Video:} a 3D VAE similar to that of \citet{wan2025wanopenadvancedlargescale}. Although it supports variable resolutions and frame rates, we normalize all videos to 540p at 24 FPS, yielding 48-channel latents of 60$\times$34 at 6 FPS.
    \item \textbf{Images:} treated as single-frame video, yielding one 48-channel 60$\times$34 latent frame for a 540p input.
\end{itemize}
Following Qwen-Image \citep{wu2025qwen}, text, image, and video inputs are additionally encoded into semantic embeddings by a pre-trained Qwen3-VL \citep{bai2025qwen3}. The backbone is therefore conditioned on clean Qwen3-VL embeddings and clean VAE latents of the input, and denoises the VAE latents of the output. It is a DiT \citep{peebles2023scalable} with MoE layers \citep{moe}, and we use Complete-muE \citep{peng2026completemue} to transfer hyperparameters from small-scale experiments.

\section{Training Details} \label{appendix:training}

\subsection{Audio-Visual Masked Reconstruction} \label{appendix:masking}
We frame AV masked reconstruction as a self-supervised AV2AV editing task (Figure~\ref{fig:masking}). Each AV sample is assigned one of three masking schemes at random:
\begin{itemize}
    \item \textbf{Temporal:} a random span of video frames is masked, and with 50\% probability the corresponding audio is zeroed out.
    \item \textbf{Spatial:} we sample $s \sim \mathcal{U}\{0, S\}$ and apply $s$ masks of 200$\times$200px to all video frames. We found $S{=}12$ to work well for our 540p videos.
    \item \textbf{Spatio-temporal:} both spatial and temporal masking are applied.
\end{itemize}
All three schemes use the same instruction (``inpaint this video''). The model is conditioned on the masked clip as its input $Y$ and must generate the clean, unmasked clip as its output $\hat{Y}$.

\subsection{Pre-Training and Supervised Fine-Tuning} \label{appendix:sft}
Because editing data is scarce, we start from an internal DiT checkpoint pre-trained on large-scale generation and editing data, which we call \textbf{CrossEdit-Base}. Its pre-training tasks are:
\begin{itemize}
    \item \textbf{Text-to-X:} text-to-image, text-to-video, and text-to-audio (text-to-speech and text-to-sound effects).
    \item \textbf{Conditional generation:} image-to-video, image-to-AV (talking head generation and animation), audio-to-video, audio-to-AV, and image+audio-to-video (lip-synced talking head generation).
    \item \textbf{Editing:} image-to-image, video-to-video, audio-to-audio (voice cloning only), and AV-to-AV (lip-synced speech editing with the mined pairs of Section~\ref{sec: mining}).
\end{itemize}
We then add the synthetic audio editing (Section~\ref{sec:synthesis}) and AV masked reconstruction (Appendix~\ref{appendix:masking}) data and fine-tune for 10K steps to obtain \textbf{CrossEdit}. SFT uses a constant learning rate of 1.2e-4, with AdamW \citep{adamw} optimizer states initialized from the pre-trained checkpoint, and takes around 2.5 days on 64 H100 GPUs or 5 days on 64 A100 GPUs. The data mixture is shown in Table~\ref{tab:sft_data_appendix}.

\subsection{Classifier-Free Guidance and Inference} \label{appendix:cfg}
During training, we independently replace the input $Y$ and the instruction $X$ with a null token $\varnothing$ at random (Section~\ref{sec:arch}). This enables classifier-free guidance (CFG)~\citep{ho2021classifierfree} with separate guidance scales for fidelity to the input ($\lambda_Y$) and adherence to the instruction ($\lambda_X$). The guided velocity is
\begin{equation}
\begin{aligned}
    \tilde{v}_\theta(\hat{Y}_t, t, Y, X) = v_\theta(\hat{Y}_t, t, \varnothing, \varnothing)
    &+ \lambda_Y \left( v_\theta(\hat{Y}_t, t, Y, \varnothing) - v_\theta(\hat{Y}_t, t, \varnothing, \varnothing) \right) \\
    &+ \lambda_X \left( v_\theta(\hat{Y}_t, t, Y, X) - v_\theta(\hat{Y}_t, t, Y, \varnothing) \right),
\end{aligned}
    \label{eq:cfg}
\end{equation}
which replaces $v_\theta$ when integrating from $t=1$ to $t=0$. We use 50 sampling steps with $\lambda_Y = \lambda_X = 5.0$ for video and $\lambda_Y = \lambda_X = 3.0$ for audio unless stated otherwise.

\begin{table}[t]
    \centering
        \caption{\textbf{SFT data mixture by total tokens and samples per batch.} Video-to-video and lip-synced speech editing samples are replaced by AV masked reconstruction (Section~\ref{sec: inpainting}) with 50\% probability.}
    \begin{tabular}{l|rr}
    \toprule
    Task  & Tokens &  Batch Samples\\
    \midrule
    \textit{Audio} \\
    Synthesized Audio Editing & 0.5B & 64 \\
    Simulated Audio Editing & 2.5B & 208 \\
    Text-to-Audio & 0.3B & 48 \\
    Speech Editing & 1.5B & 96 \\
    \midrule
    \textit{Video} \\
    Text to Video & 1.3B & 4 \\
    Video to Video & 1.3B & 8 \\
    Image+Audio to Video & 3.2B & 10 \\
    \midrule
    \textit{Audio-Visual} \\
    Text to AV & 1.3B & 4 \\
    Lip-Synced Speech Editing & 2.0B & 8 \\
    \midrule
    \textit{Image} \\
    Image to Image & 0.2B & 12 \\
    \bottomrule
    \end{tabular}
    \label{tab:sft_data_appendix}
\end{table}

\section{Evaluation Details} \label{appendix:evaluation}

\subsection{CrossEditBench} \label{appendix:benchmark}
\begin{figure}
    \centering
    \includegraphics[width=1\linewidth]{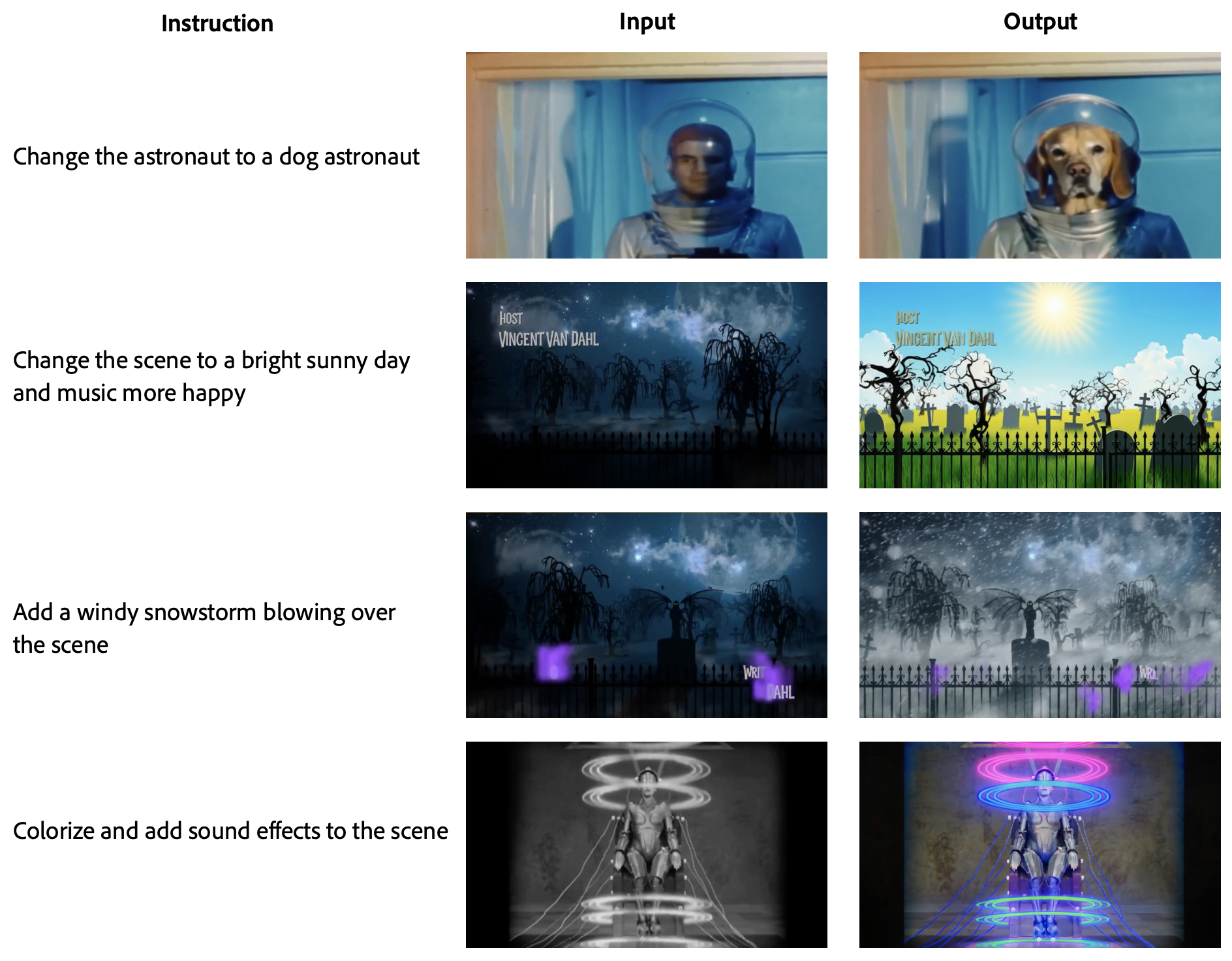}
    \caption{Example AV editing instructions, inputs, and outputs for CrossEditBench/CrossEdit.}
    \label{fig:examples_1}
\end{figure}

\begin{figure}
    \centering
    \includegraphics[width=1\linewidth]{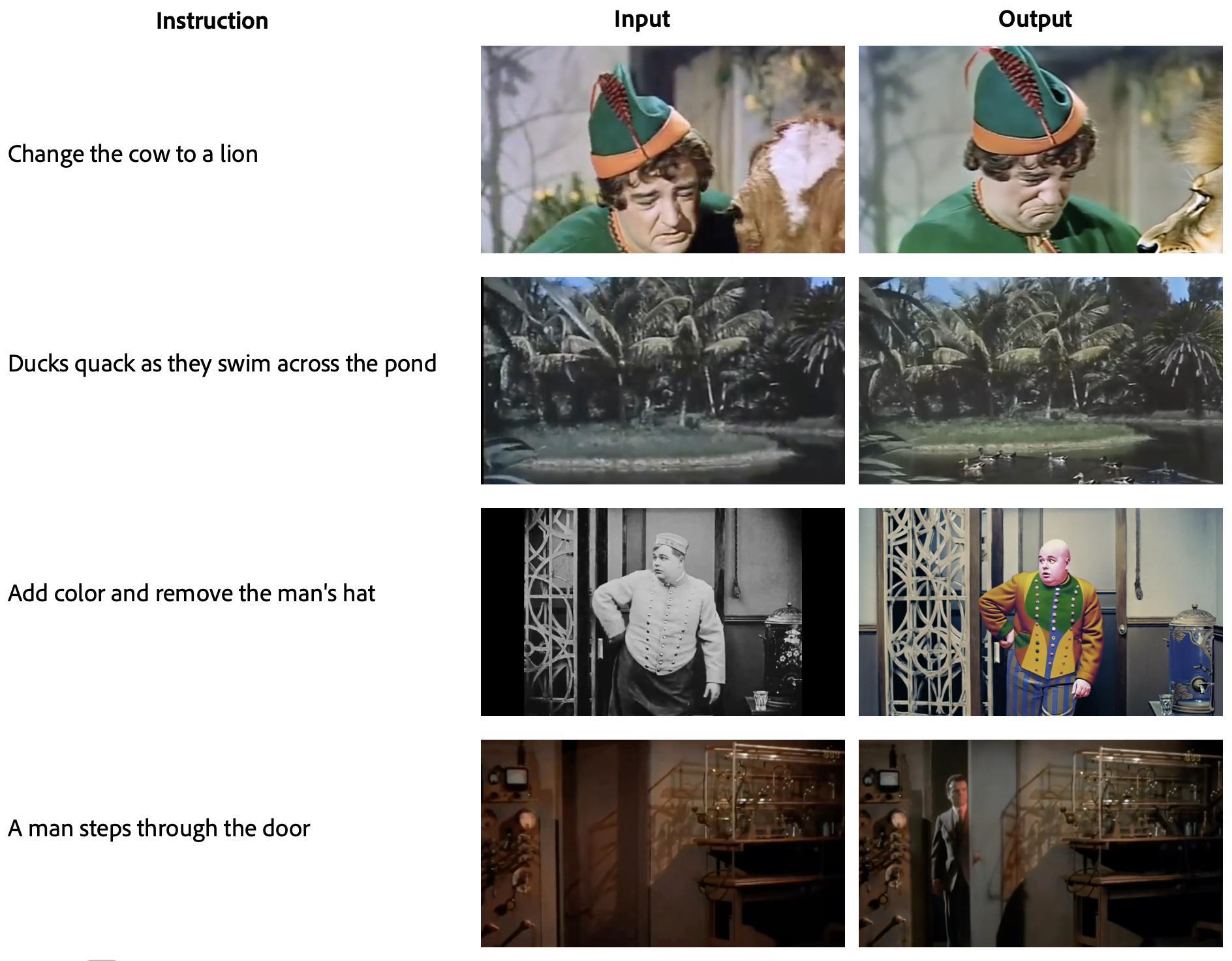}
    \caption{Example AV editing instructions, inputs, and outputs for CrossEditBench/CrossEdit.}
    \label{fig:examples_2}
\end{figure}

CrossEditBench evaluates editing of videos with audio on realistic use cases. We annotate 100 clips from OSSL \citep{ossl}, an open-source dataset of public-domain movies originally built for music processing. Each annotation contains the edit instruction (``add a cow and remove the music''), what should change in the audio ([added: cow mooing, removed: music]), and what should change in the video ([added: cow]). We do not use the per-modality annotations in this work but release them to support fine-grained evaluation. As described in Section~\ref{sec:benchmark}, instructions vary in which modality they target and whether they require AV synchrony: an edit may affect both modalities (adding a barking dog), only one modality but in sync with the other (adding a person walking in time with audible footsteps), or only one modality without synchrony (removing the music). Benchmark examples are shown in Figure~\ref{fig:demo}, with additional inputs and outputs in Figures~\ref{fig:examples_1} and~\ref{fig:examples_2}.

\subsection{Subjective Evaluation} \label{appendix: subjective}
\begin{figure}
    \centering
    \includegraphics[width=1\linewidth]{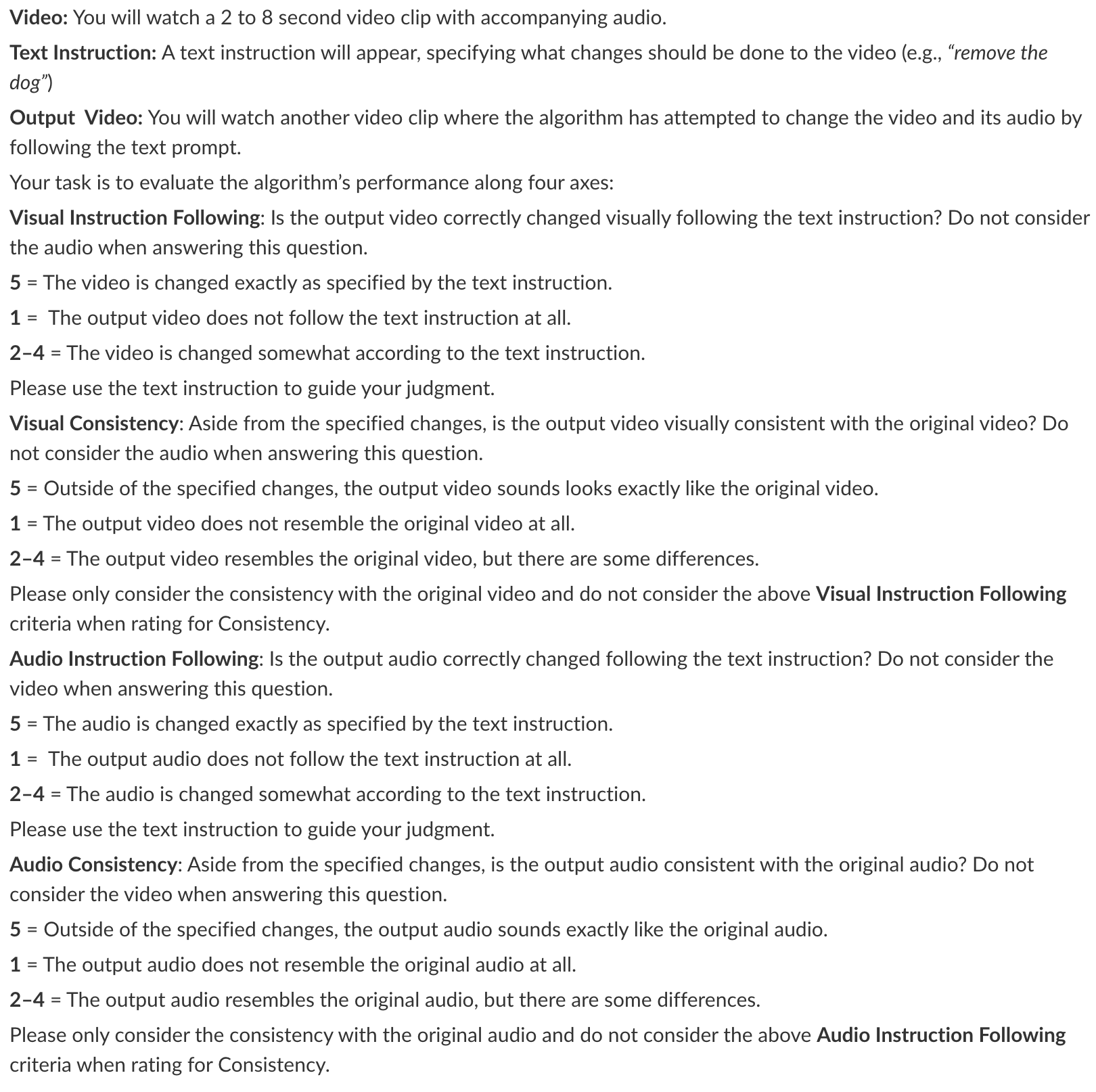}
    \caption{Instructions given to human annotators and the LLM Judges}
    \label{fig:subjective}
\end{figure}

For AV scene editing, we recruited 100 crowdsourced raters, restricted to English speakers with at least 100 completed tasks and a task approval rate of at least 98\%, and required them to wear headphones. Each rater rated 4 clips per system on a 1--5 scale (higher is better) along four axes: visual instruction following, visual consistency, audio instruction following, and audio consistency (instructions in Figure~\ref{fig:subjective}). The unedited input is included as a control system to validate rater attention: we exclude the 14 raters who gave it low consistency scores, leaving 86 raters and 344 ratings per system.
Full results with 95\% confidence intervals, including the + Audio Editing ablation for both tasks, are in Table~\ref{tab:data_ablation} (AV scene editing) and Table~\ref{tab:data_ablation_appendix_speech} (lip-synced speech editing).

The study was a minimal-risk human evaluation task in which participants viewed videos and provided quality ratings. No sensitive personal information was collected, and participants were informed about the nature of the task and the use of their responses for research purposes. Participation was voluntary, and participants were compensated at approximately \$15 USD per hour. Under our institution's policies, this type of study does not require IRB review or approval.

\begin{table}[h]
    \centering
    \caption{\textbf{Full subjective results on IMDB lip-synced speech editing} (1--5, mean $\pm$ 95\% confidence interval).}
    \resizebox{\textwidth}{!}{
    \begin{tabular}{l|cccc}
    \toprule
     & \multicolumn{2}{c}{Audio}  & \multicolumn{2}{c}{Video} \\
     & Ins. & Cons. & Ins. & Cons.  \\
     \midrule
    Input & 3.29$\pm$0.25 & 4.39$\pm$0.18 & 3.29$\pm$0.25 & 4.50$\pm$0.16 \\
    \midrule
    F5-TTS+HuMo & 3.30$\pm$0.23 & 2.97$\pm$0.22 & 2.60$\pm$0.22 & 2.10$\pm$0.18 \\
    F5-TTS+StableAvatar & 3.21$\pm$0.23 & 2.88$\pm$0.22 & 2.17$\pm$0.22 & 1.78$\pm$0.20 \\
    \midrule
    CrossEdit-Base & \textbf{3.98}$\pm$0.19 & 3.81$\pm$0.19 & \textbf{4.05}$\pm$0.18 & 4.22$\pm$0.17  \\
    \hspace{0.3cm} + Audio Editing  &  3.86$\pm$0.20 & \textbf{3.82}$\pm$0.20 & 3.99$\pm$0.19 & 4.35$\pm$0.16 \\
        \hspace{0.7cm}+ AV Masked Reconstruction (CrossEdit) & 3.96$\pm$0.20 & \textbf{3.82}$\pm$0.20 & 4.02$\pm$0.19 & \textbf{4.42}$\pm$0.15\\
    \bottomrule
    \end{tabular}}
    \label{tab:data_ablation_appendix_speech}
\end{table}

\subsection{AV-FES Details} \label{appendix:avfes}
\paragraph{Motivation.} Instruction following and consistency cannot individually measure editing performance well. A system that returns its input untouched scores perfectly on consistency, and a system that regenerates the whole clip can score well on instruction following while discarding everything the instruction did not mention. Neither axis alone answers the question we care about: \emph{did the system make the requested change and nothing else?} Even when both scores are available, it is difficult to weigh increases in one score against drops in the other. AV-FES (Section~\ref{sec:avfes}) answers it with a single score.

\paragraph{Zero headroom.} In the rare case that $\mathrm{Ins}_m(Y) = 5$, the judge deems the instruction already satisfied for modality $m$, and the gain in Equation~\ref{eq:gain} is undefined. The correct action is then to leave the modality alone, so we set $g_m(\hat{Y}) = \mathbf{1}\!\left[\mathrm{Cons}_m(\hat{Y}) \geq 4\right]$.

\paragraph{Why the harmonic mean.} As in the $F_1$ score, $g_m$ plays the role of recall (how much of the requested change was made) and $c_m$ that of precision (how much of the output is free of unrequested change). The harmonic mean in Equation~\ref{eq:fes} is high only when \emph{both} are high: an untouched copy scores $0$ because $g_m = 0$, and a full re-synthesis that happens to satisfy the instruction is pulled down by its low $c_m$. When both terms vanish, we set $\text{FES}_m(\hat{Y}) = 0$.

\paragraph{Range and ceiling.} AV-FES is bounded in $[0,1]$, is zero for the unedited input by construction, and uses the same judge ratings as the Ins.\ and Cons.\ subscores in Table~\ref{tab:scene_edit}. In practice, the Gemini judge awards at most 4 on consistency even for byte-identical clips (cf.\ the input's consistency scores in Table~\ref{tab:scene_edit}), so $c_m$ saturates near $0.75$ and the empirical ceiling of AV-FES for a flawless editor is about $0.86$ rather than $1$. This compresses the scale uniformly and does not affect rankings.

\paragraph{Single-modality instructions.} When an instruction concerns only one modality, the correct behavior in the other modality is to leave it unchanged. AV-FES therefore rewards edits where they were requested and, through $c_m$, penalizes collateral changes where they were not.

\subsection{Robustness to the Choice of Judge} \label{appendix:judge}
To test whether our conclusions depend on the LLM judge, we re-score CrossEditBench outputs with Qwen3-Omni \citep{xu2025qwen3} using the same rubric and protocol as the Gemini 3.1 Pro results in Table~\ref{tab:scene_edit} (100 clips per system). Results for the systems in this paper are in Table~\ref{tab:qwen_judge}. To measure agreement between the judges, we compute system-level correlations over a larger pool of 11 systems, including development checkpoints not reported elsewhere in the paper and the unedited input (Table~\ref{tab:judge_corr}). Because the input scores 0 under both judges by construction, we also report correlations without it (10 systems). The judges agree strongly on system rankings in both cases (AV-FES Pearson 0.87, Spearman 0.80 without the input), and both rank CrossEdit first and CrossEdit-Base second among the reported systems. Agreement is highest for video and lower for audio. Qwen3-Omni is considerably more lenient, however: its scores span a range roughly two to three times wider than Gemini's, and it awards full consistency to the unedited input. AV-FES values should therefore be compared only within a single judge, not across judges.

We also compare each judge against the human ratings of Table~\ref{tab:data_ablation}, which cover a subset of systems: 4 shared with Gemini (the two LucyEdit cascades, CrossEdit-Base, and CrossEdit) and 5 shared with Qwen3-Omni (additionally the + Audio Editing ablation). Both judges rank systems much as humans do (Table~\ref{tab:human_judge_corr}), with the largest disagreement on Qwen3-Omni's visual consistency, where it rates CrossEdit above the cascades while humans do not. With so few systems, these correlations are coarse, and a single system can change them substantially.

\subsection{Confidence Intervals for LLM-Judge Scores} \label{appendix:judge_ci}
Tables~\ref{tab:gemini_ci} and~\ref{tab:qwen_ci} report 95\% confidence intervals over the 100 CrossEditBench clips for both judges. CrossEdit's AV-FES exceeds that of every baseline under both judges; its margins over the cascades, AvED, AVI-EDIT, and InstructAV2AV are well outside the intervals, while its margin over MiniMax H3 is smaller but consistent across judges. 

\begin{table}[h]
    \centering
    \caption{\textbf{CrossEditBench results with Gemini 3.1 Pro as the judge, with 95\% confidence intervals} (mean $\pm$ half-width over the 100 clips). Means match Table~\ref{tab:scene_edit}.}
    \resizebox{\textwidth}{!}{
    \begin{tabular}{l|cccc|ccc}
    \toprule
     & \multicolumn{4}{c|}{Judge axes (1--5)} & \multicolumn{3}{c}{Faithful Edit Score} \\
     & Vis.\ Ins. & Vis.\ Cons. & Aud.\ Ins. & Aud.\ Cons. & V-FES $\uparrow$ & A-FES $\uparrow$ & \textbf{AV-FES} $\uparrow$ \\
    \midrule
    Input              & 2.81$\pm$0.19 & 3.97$\pm$0.03 & 2.19$\pm$0.10 & 3.95$\pm$0.05 & -- & -- & -- \\
    \midrule
    LucyEdit + HY-XL   & 2.88$\pm$0.15 & 3.76$\pm$0.16 & 2.79$\pm$0.18 & 2.27$\pm$0.07 & 0.12$\pm$0.02 & 0.19$\pm$0.04 & 0.16$\pm$0.03 \\
    LucyEdit + HY-XXL  & 2.88$\pm$0.16 & 3.76$\pm$0.16 & 2.89$\pm$0.17 & 2.15$\pm$0.10 & 0.12$\pm$0.04 & 0.20$\pm$0.05 & 0.16$\pm$0.03 \\
    VACE + Coherent    & 3.16$\pm$0.17 & 3.44$\pm$0.16 & 3.12$\pm$0.17 & 2.49$\pm$0.14 & 0.12$\pm$0.04 & 0.26$\pm$0.04 & 0.19$\pm$0.03 \\
    AvED               & 2.81$\pm$0.16 & 3.20$\pm$0.13 & 2.49$\pm$0.17 & 3.93$\pm$0.05 & 0.09$\pm$0.04 & 0.14$\pm$0.05 & 0.12$\pm$0.03 \\
    AVI-EDIT           & 2.42$\pm$0.12 & 2.35$\pm$0.12 & 2.93$\pm$0.18 & 3.65$\pm$0.14 & 0.08$\pm$0.03 & 0.25$\pm$0.06 & 0.17$\pm$0.03 \\
    \midrule
    InstructAV2AV      & 2.85$\pm$0.17 & 3.03$\pm$0.17 & 2.85$\pm$0.18 & 3.20$\pm$0.17 & 0.15$\pm$0.05 & 0.24$\pm$0.05 & 0.20$\pm$0.04 \\
    MiniMax H3         & 3.14$\pm$0.16 & 3.38$\pm$0.15 & 2.77$\pm$0.18 & 3.70$\pm$0.11 & 0.24$\pm$0.06 & 0.22$\pm$0.06 & 0.23$\pm$0.04 \\
    \midrule
    CrossEdit-Base     & 2.88$\pm$0.14 & 2.00$\pm$0.12 & 3.15$\pm$0.18 & 3.52$\pm$0.18 & 0.20$\pm$0.04 & 0.31$\pm$0.05 & 0.26$\pm$0.04 \\
    CrossEdit          & 3.59$\pm$0.13 & 2.81$\pm$0.16 & 3.70$\pm$0.18 & 3.84$\pm$0.17 & 0.26$\pm$0.06 & 0.33$\pm$0.06 & 0.30$\pm$0.04 \\
    \bottomrule
    \end{tabular}}
    \label{tab:gemini_ci}
\end{table}

\begin{table}[h]
    \centering
    \caption{\textbf{CrossEditBench results with Qwen3-Omni as the judge, with 95\% confidence intervals} (mean $\pm$ half-width over the 100 clips). Means match Table~\ref{tab:qwen_judge}.}
    \resizebox{\textwidth}{!}{
    \begin{tabular}{l|cccc|ccc}
    \toprule
     & \multicolumn{4}{c|}{Judge axes (1--5)} & \multicolumn{3}{c}{Faithful Edit Score} \\
     & Vis.\ Ins. & Vis.\ Cons. & Aud.\ Ins. & Aud.\ Cons. & V-FES $\uparrow$ & A-FES $\uparrow$ & \textbf{AV-FES} $\uparrow$ \\
    \midrule
    Input              & 2.69$\pm$0.39 & 5.00$\pm$0.00 & 3.32$\pm$0.40 & 4.40$\pm$0.28 & -- & -- & -- \\
    \midrule
    LucyEdit + HY-XL   & 2.68$\pm$0.38 & 4.10$\pm$0.29 & 3.44$\pm$0.38 & 2.47$\pm$0.36 & 0.16$\pm$0.09 & 0.19$\pm$0.12 & 0.17$\pm$0.07 \\
    LucyEdit + HY-XXL  & 3.88$\pm$0.35 & 4.18$\pm$0.25 & 3.84$\pm$0.34 & 2.55$\pm$0.38 & 0.50$\pm$0.11 & 0.28$\pm$0.13 & 0.41$\pm$0.09 \\
    VACE + Coherent    & 3.85$\pm$0.34 & 4.20$\pm$0.24 & 3.73$\pm$0.38 & 3.03$\pm$0.38 & 0.33$\pm$0.10 & 0.15$\pm$0.10 & 0.25$\pm$0.07 \\
    AvED               & 3.70$\pm$0.37 & 4.48$\pm$0.18 & 3.68$\pm$0.36 & 4.23$\pm$0.30 & 0.46$\pm$0.12 & 0.26$\pm$0.13 & 0.37$\pm$0.09 \\
    AVI-EDIT           & 3.19$\pm$0.39 & 2.59$\pm$0.33 & 4.32$\pm$0.30 & 3.02$\pm$0.39 & 0.39$\pm$0.10 & 0.33$\pm$0.13 & 0.36$\pm$0.08 \\
    \midrule
    InstructAV2AV      & 3.77$\pm$0.36 & 4.25$\pm$0.21 & 4.44$\pm$0.28 & 3.79$\pm$0.35 & 0.54$\pm$0.12 & 0.58$\pm$0.15 & 0.55$\pm$0.09 \\
    MiniMax H3         & 4.19$\pm$0.30 & 4.68$\pm$0.14 & 4.13$\pm$0.32 & 4.13$\pm$0.33 & 0.69$\pm$0.11 & 0.55$\pm$0.14 & 0.63$\pm$0.09 \\
    \midrule
    CrossEdit-Base     & 4.16$\pm$0.32 & 3.10$\pm$0.29 & 4.60$\pm$0.24 & 3.03$\pm$0.36 & 0.55$\pm$0.09 & 0.41$\pm$0.13 & 0.49$\pm$0.08 \\
    CrossEdit          & 4.61$\pm$0.23 & 4.44$\pm$0.20 & 4.72$\pm$0.20 & 4.36$\pm$0.27 & 0.80$\pm$0.08 & 0.74$\pm$0.12 & 0.77$\pm$0.07 \\
    \bottomrule
    \end{tabular}}
    \label{tab:qwen_ci}
\end{table}

\begin{table}[h]
    \centering
    \caption{\textbf{CrossEditBench results with Qwen3-Omni as the judge.} Same rubric and protocol as Table~\ref{tab:scene_edit}. AV-FES$|$edit averages FES$_m$ only over clips whose instruction targets modality $m$. No-gain: number of clips (out of 100) with zero instruction gain.}
    \resizebox{\textwidth}{!}{
    \begin{tabular}{l|cccc|ccc|cc}
    \toprule
     & \multicolumn{4}{c|}{Judge axes (1--5)} & & & & & \\
     & Vis. Ins. & Vis. Cons. & Aud. Ins. & Aud. Cons. & V-FES $\uparrow$ & A-FES $\uparrow$ & \textbf{AV-FES} $\uparrow$ & AV-FES$|$edit $\uparrow$ & No-gain $\downarrow$ \\
    \midrule
    Input & 2.69 & 5.00 & 3.32 & 4.40 & -- & -- & -- & -- & -- \\
    \midrule
    LucyEdit + HY-XL  & 2.68 & 4.10 & 3.44 & 2.47 & 0.16 & 0.19 & 0.17 & 0.17 & 75 \\
    LucyEdit + HY-XXL & 3.88 & 4.18 & 3.84 & 2.55 & 0.50 & 0.28 & 0.41 & 0.38 & 55 \\
    VACE + Coherent   & 3.85 & 4.20 & 3.73 & 3.03 & 0.33 & 0.15 & 0.25 & 0.23 & 60 \\
    AvED              & 3.70 & 4.48 & 3.68 & 4.23 & 0.46 & 0.26 & 0.37 & 0.35 & 62 \\
    AVI-EDIT          & 3.19 & 2.59 & 4.32 & 3.02 & 0.39 & 0.33 & 0.36 & 0.40 & 53 \\
    MiniMax-H3        & 4.19 & \textbf{4.68} & 4.13 & 4.13 & 0.69 & 0.55 & 0.63 & 0.63 & 48 \\
    InstructAV2AV     & 3.77 & 4.25 & 4.44 & 3.79 & 0.54 & 0.58 & 0.55 & 0.56 & 50 \\
    \midrule
    CrossEdit-Base    & 4.16 & 3.10 & 4.60 & 3.03 & 0.55 & 0.41 & 0.49 & 0.50 & 38 \\
    CrossEdit         & \textbf{4.61} & 4.44 & \textbf{4.72} & \textbf{4.36} & \textbf{0.80} & \textbf{0.74} & \textbf{0.77} & \textbf{0.75} & \textbf{34} \\
    \bottomrule
    \end{tabular}}
    \label{tab:qwen_judge}
\end{table}

\begin{table}[h]
    \centering
    \caption{\textbf{System-level agreement between Gemini 3.1 Pro and Qwen3-Omni judges.} Computed over 13 systems, including development checkpoints not reported in the main paper and the unedited input, which scores 0 under both judges by construction. The last column excludes the input (12 systems). AV-FES$|$edit averages FES$_m$ only over clips whose instruction targets modality $m$. Mean $|$diff$|$ is in score units.}
    \resizebox{\textwidth}{!}{
    \begin{tabular}{l|ccc|cc|c}
    \toprule
     & \multicolumn{5}{c|}{All systems ($n{=}13$)} & w/o input ($n{=}12$) \\
    Score & Pearson & Spearman & Mean $|$diff$|$ & Gemini range & Qwen range & Pearson / Spearman \\
    \midrule
    V-FES           & 0.96 & 0.95 & 0.356 & 0.00--0.29 & 0.00--0.80 & 0.95 / 0.93 \\
    A-FES           & 0.74 & 0.70 & 0.185 & 0.00--0.38 & 0.00--0.74 & 0.60 / 0.62 \\
    \midrule
    \textbf{AV-FES} & \textbf{0.90} & \textbf{0.90} & 0.270 & 0.00--0.34 & 0.00--0.77 & \textbf{0.85 / 0.87} \\
    AV-FES$|$edit   & 0.89 & 0.87 & 0.215 & 0.00--0.45 & 0.00--0.75 & 0.83 / 0.83 \\
    \bottomrule
    \end{tabular}}
    \label{tab:judge_corr}
\end{table}

\begin{table}[h]
    \centering
    \caption{\textbf{System-level agreement between LLM judges and human ratings} (Pearson / Spearman) over the systems shared with Table~\ref{tab:data_ablation}: $n{=}4$ for Gemini 3.1 Pro and $n{=}5$ for Qwen3-Omni.}
    \begin{tabular}{l|cc}
    \toprule
    Score & Gemini 3.1 Pro & Qwen3-Omni \\
    \midrule
    Vis. Ins.   & 0.93 / 1.00 & 0.82 / 1.00 \\
    Vis. Cons.  & 0.99 / 1.00 & 0.71 / 0.60 \\
    Aud. Ins.   & 0.71 / 0.50 & 0.94 / 0.97 \\
    Aud. Cons.  & 1.00 / 0.90 & 0.93 / 1.00 \\
    \midrule
    V-FES       & 0.92 / 1.00 & 0.90 / 1.00 \\
    A-FES       & 0.84 / 0.50 & 0.99 / 1.00 \\
    \textbf{AV-FES} & \textbf{0.92 / 1.00} & \textbf{0.96 / 1.00} \\
    AV-FES$|$edit & 0.94 / 0.90 & 0.97 / 1.00 \\
    \bottomrule
    \end{tabular}
    \label{tab:human_judge_corr}
\end{table}

\subsection{Lip-Synced Speech Editing} \label{appendix:speech_edit}
Lip-synced speech editing is the task of generating new speech and matching lip movements from an instruction and a reference AV clip; it is the audio-visual counterpart of voice cloning (Figure~\ref{fig:arch}). Our IMDB test set consists of 600 two-second clips of different celebrities talking. The videos are shot in varied environments and often include challenging conditions such as multiple visible people (with only one speaker), moving arms and hands, and camera panning. Example inputs and outputs are on our demo page.

\begin{figure}[h]
    \centering
    \includegraphics[width=0.75\linewidth]{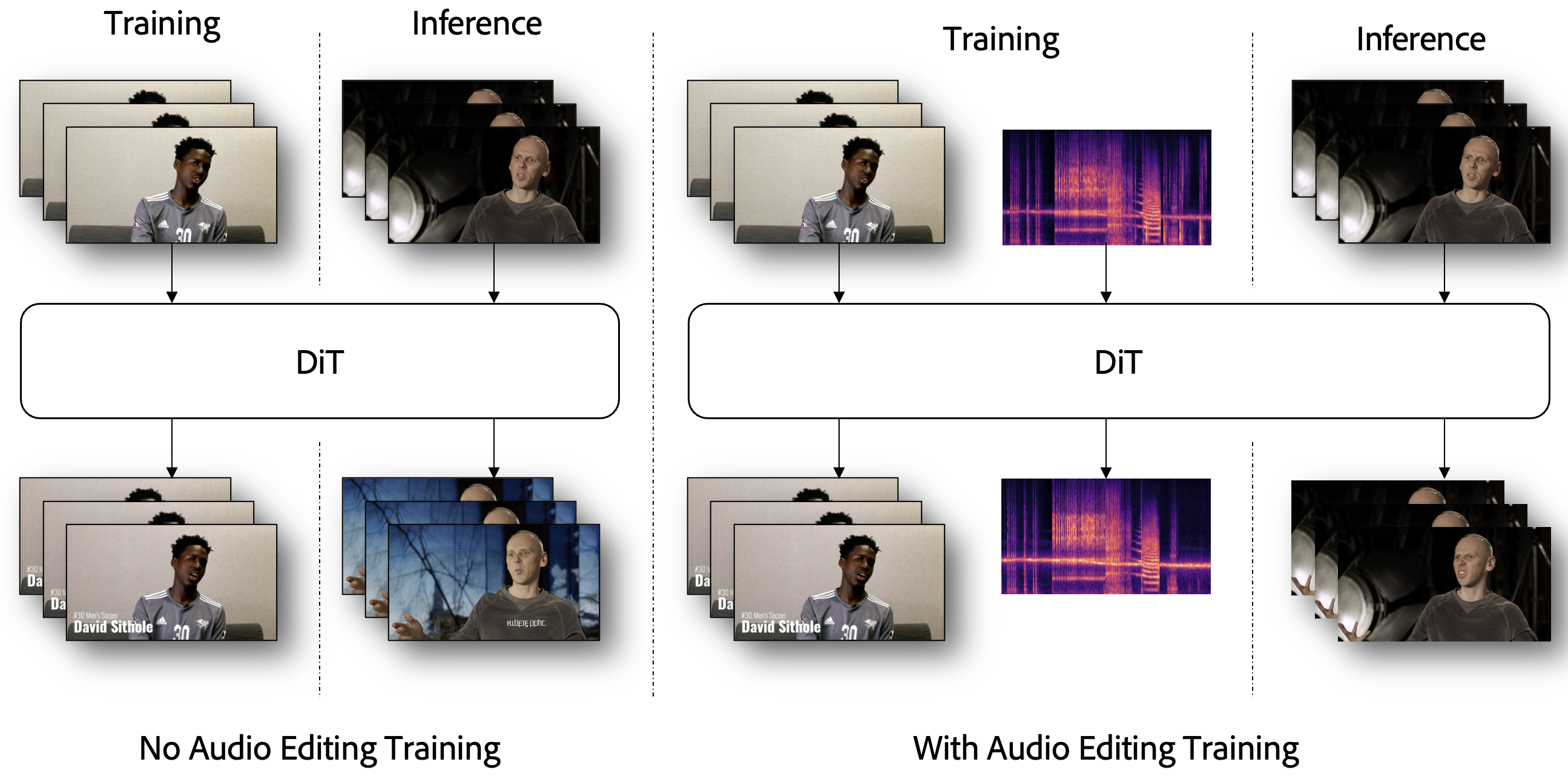}
    \caption{\textbf{Synthetic audio editing data reinforces visual consistency.} Mined AV pairs are not perfectly consistent (here, the target has a text overlay and different lighting), causing the model to hallucinate unwanted changes (left). Our method corrects this (right).}
    \label{fig:consistency}
\end{figure}

\section{Audio Editing Results} \label{sec:audio_edit}
\textbf{Setup:} We evaluate on three test sets:
\begin{itemize}
    \item \textbf{StoryGenEval} \citep{chen2026audiochat}: complex synthetic scenes generated by a tool-calling LLM agent (similar to WavJourney \citep{wavjourney}) that uses generative audio models as tools. Instructions are free-form and open-vocabulary.
    \item \textbf{AudioCapsEdit} \citep{chen2026audiochat}: real-world scenes from AudioCaps \citep{kim-etal-2019-audiocaps} with synthesized edit instructions, testing whether models trained on synthetic data generalize to real recordings. There is no ground-truth output.
    \item \textbf{Movie Separation}: an internal set of professionally mixed movie trailers for instruction-based source separation. The model must extract or remove a list of categories (ambience, music, dialogue, and sound effects). There is no ground-truth output.
\end{itemize}
Our primary metrics are $\Delta$FLAM (consistency with the input) and editFLAM (instruction following), which \citet{chen2026audiochat} show correlate better with human judgment than FAD \citep{fad} and CLAPScore \citep{wu2023large}. On Movie Separation, we also report SAM-Audio-Judge \citep{wang2026samaudiojudgeunified}, averaged over judging categories. We compare to AudioChat \citep{chen2026audiochat}, a SOTA instruction-based audio editing model, and SAM-Audio Large \citep{shi2025samaudiosegmentaudio}, a SOTA instruction-based source separation model. Since SAM-Audio only supports sound removal, we invert extraction instructions before passing them to it (``extract the dialogue'' $\rightarrow$ ``remove the music, ambience, and sound effects''). CrossEdit-Base is omitted as it was not trained for complex audio editing.

\textbf{Results:} CrossEdit outperforms prior work on all three test sets (Table~\ref{tab:audio_editing}). The largest gain is on AudioCapsEdit (40.3 vs.\ 15.5 editFLAM over AudioChat), which we attribute to our large-scale synthetic mixtures. CrossEdit also surpasses SAM-Audio on source separation, suggesting that our synthetic data strategy is more effective than its pseudo-labeling of real mixtures.

\begin{table}[t]
    \centering
    \caption{\textbf{Audio editing results.} StoryGenEval tests complex instruction following, AudioCapsEdit real-world editing, and Movie Separation instruction-based source separation. SAM-J: SAM-Audio-Judge.}
    \resizebox{\textwidth}{!}{
    \begin{tabular}{l|cc|c|ccc}
    \toprule
     & \multicolumn{2}{c|}{StoryGenEval} & \multicolumn{1}{c|}{AudioCapsEdit} & \multicolumn{3}{c}{Movie Separation}  \\
     & $\Delta$FLAM $\downarrow$ & editFLAM $\uparrow$ & editFLAM $\uparrow$ & $\Delta$FLAM $\downarrow$ & editFLAM $\uparrow$ & SAM-J $\uparrow$\\
     \midrule
     SAM-Audio & - & - & - & 18.97 & 4.12 & 2.96 \\
     AudioChat  & 11.7 & 18.6 & 15.5 & - & - & - \\
     \midrule
     CrossEdit & \textbf{6.64} & \textbf{23.1} & \textbf{40.3} & \textbf{7.08} & \textbf{4.29} & \textbf{3.41}\\
     \bottomrule
    \end{tabular}}
    \label{tab:audio_editing}
\end{table}

\end{document}